\documentclass[12pt]{article}

\usepackage{amssymb}
\usepackage{amsmath}
\usepackage{amscd}
\usepackage{latexsym}
\usepackage{graphicx}
\usepackage{url}

\usepackage{cite}
\usepackage{caption}

\newcommand{\be}{\begin{equation}}
\newcommand{\ee}{\end{equation}}
\newcommand{\Dlt}{\Delta}
\newcommand{\dlt}{\delta}

\newcommand{\br}{{\bf r}}
\newcommand{\bk}{{\bf k}}
\newcommand{\bbe}{{\bf e}}

\newcommand{\bbn}{{\bf n}}
\newcommand{\bt}{\beta}
\newcommand{\vp}{\varphi}

\newcommand{\al}{\alpha}
\newcommand{\ra}{\rightarrow}

\newcommand{\gm}{\gamma}
\newcommand{\om}{\omega}
\newcommand{\Om}{\Omega}
\newcommand{\Gm}{\Gamma}

\newcommand{\lbd}{\lambda}

\newcommand{\bB}{{\bf B}}

\newcommand{\bS}{{\bf S}}

\newcommand{\rgl}{\rangle}
\newcommand{\lgl}{\langle}

\begin{document}

\begin{center}

{\Large{\bf Coherent radiation by magnets with exchange 
interactions} \\ [5mm]

V.I. Yukalov$^{1,*}$ and E.P. Yukalova$^2$ } \\ [3mm]

{\it
$^1$Bogolubov Laboratory of Theoretical Physics, \\
Joint Institute for Nuclear Research, Dubna 141980, Russia \\ [3mm]

$^2$Laboratory of Information Technologies, \\
Joint Institute for Nuclear Research, Dubna 141980, Russia }

\end{center}

\vskip 3cm

\begin{abstract}

A wide class of materials acquires magnetic properties due to particle 
interactions through exchange forces. These can be atoms and molecules 
composing the system itself, as in the case of numerous magnetic substances.
Or these could be different defects, as in the case of graphene, graphite, 
carbon nanotubes, and related materials. The theory is suggested 
describing fast magnetization reversal in magnetic systems, whose 
magnetism is caused by exchange interactions. The effect is based on the 
coupling of a magnetic sample with an electric circuit producing a feedback 
magnetic field. This method can find various applications in spintronics.
The magnetization reversal can be self-organized, producing spin 
superradiance. A part of radiation is absorbed by a resonator magnetic 
coil. But an essential part of radiation can also be emitted through the 
coil sides.

\end{abstract}

\vskip 1cm

{\parindent=0pt
{\bf Keywords}: Magnetic materials, Exchange interactions, Magnetic graphene,
Spintronics, Spin superradiance, Maser radiation   

\vskip 1cm

{\bf PACS numbers}: 84.40.Dc; 84.40.Ik; 84.90+a; 85.75.Hh; 85.75.Ff

\vskip 2cm

{\bf $^*$}Corresponding author: V.I. Yukalov

{\bf E-mail}: yukalov@theor.jinr.ru

}

\newpage

\section{Introduction}

Magnetic materials, whose magnetism is caused by exchange interactions, 
form a very wide class of magnets of different sizes, including the samples
of nanosizes, which find a variety of applications \cite{Evans_1}. A novel 
class of exchange-interaction magnetic materials is the graphene family with 
defects, including graphene flakes and ribbons, graphite, and carbon 
nanotubes \cite{Yaziev_2,Katsnelson_3,Enoki_4}. Magnetic materials find 
numerous applications in quantum electronics, for example, in spintronics, 
information processing etc. 

In the present paper, we concentrate on two interconnected features of
magnetic materials: (i) First, we study the way of fast magnetization 
reversal in these materials that is a property crucially important for 
spintronics and information processing. We show that by coupling a magnetic 
sample to a resonant electric circuit it is possible to achieve fast 
magnetization reversal. The details of the reversal can be easily regulated
by varying the system parameters. (ii) Second, fast magnetization reversal 
should produce magneto-dipole spin radiation in radio-frequency or microwave 
region. However, a large part of this radiation would be absorbed by the coil 
of the coupled electric circuit. We aim at studying whether some part of the 
maser radiation could be emitted through the coil sides. The possibility of
this effect would essentially widen the region of applicability of such
magnetic materials. 

The fact that coupling a resonant electric circuit to a spin system could
essentially influence spin dynamics is the essence of the Purcell effect 
\cite{Purcell_5}. A detailed theory of spin dynamics, employing the Purcell 
effect, has been developed for nuclear spins 
\cite{Yukalov_6,Yukalov_7,Yukalov_8,Yukalov_9,Yukalov_10},
magnetic nanomolecules \cite{Yukalov_11,Yukalov_12,Yukalov_13}, 
and magnetic nanoclusters \cite{Yukalov_14,Yukalov_15,Yukalov_16,Kharebov_17}
(see also the review article \cite{Yukalov_18}). The characteristic feature 
of all these materials is that their particle interactions are described by
dipolar forces. Also, only the total radiation intensity has been considered.
However, in the typical setup, the magnetic sample is inserted into a coil
of a resonant magnetic circuit, so that an essential part of radiation is 
absorbed by this coil. In order to understand whether some part of 
magneto-dipole radiation could be emitted through the coil sides, it is 
necessary to study the spatial distribution of the radiation.

The main novelty of the present paper is twofold: (i) We consider the class 
of magnetic materials with exchange interactions. This class is widespread 
and important, including a number of known magnetic materials as well as 
new nanomaterials, such as graphene with defects. (ii) We study the spatial 
distribution of radiation in order to conclude whether a sample inside a coil 
could serve as an emitter of radiation passing through the coil sides.

\section{Magnets with exchange interactions}

The Hamiltonian of a system of $N$ particles is the sum 
\be
\label{1} 
 \hat H = \hat H_{ex} - \mu_0 \sum_{j=1}^N \bB \cdot \bS_j \;  .
\ee
The first term is the Hamiltonian of an anisotropic Heisenberg model,
\be
\label{2}
 \hat H_{ex} = -\; \frac{1}{2} \sum_{i\neq j} \left [ J_{ij} \left (
S_i^x S_j^x +  S_i^y S_j^y \right ) + I_{ij} S_i^z S_j^z \right ]\;,
\ee
describing particles with exchange interactions. The second is a Zeeman term, 
with $\mu_0$ being magnetic moment, and with the total magnetic field
\be
\label{3}
 \bB = B_0 \bbe_z + H \bbe_x  
\ee
consisting of an external magnetic field $B_0$ along the $z$ axis and a 
feedback field $H$, along the $x$ axis, caused by the magnetic coil of an 
electric circuit. The sample is inserted into the coil, with its axis along
the $x$ axis. The feedback field is defined by the Kirchhoff equation that 
can be written \cite{Yukalov_8,Yukalov_9,Yukalov_18} as
\be
\label{4}
  \frac{dH}{dt} + 2\gm H + \om^2 \int_0^t H(t') \; dt' = 
- 4\pi \; \frac{dm_x}{dt} \; .
\ee
Here $\gamma$ is the circuit attenuation, $\omega$ is the circuit natural 
frequency, and the effective electromotive force is due to the moving 
magnetization of the sample,
\be
\label{5}
 m_x = \frac{\mu_0}{V_c} \; \sum_{j=1}^N \; \lgl S_j^x \rgl \;  ,
\ee
where $V_c$ is the coil volume and angle brackets imply statistical 
averaging. 

The $z$ axis is assumed to be the axis of the easy magnetization, so that
$I_{ij}$ should be larger than $J_{ij}$. If exchange interactions are due 
to electrons, then $\mu_0 < 0$. Therefore, if $B_0 > 0$, then the equilibrium 
spin value $S_0$ of a particle is negative, $S_0 < 0$. In this case, the
Zeeman frequency is positive,
\be
\label{6}
 \om_0 \equiv -\mu_0 B_0 > 0 \;  .
\ee

The effective particle spin $S$ can be arbitrary. The ladder spin operators
$$
S_j^\pm \equiv S_j^x \pm i S_j^y 
$$
satisfy the commutation relations
$$
 \left [ S_i^+ , \; S_j^- \right ] = 2\dlt_{ij} S_j^z \; , \qquad
  \left [ S_i^- , \; S_j^z \right ] = 2\dlt_{ij} S_j^- \; .
$$
In terms of the ladder spin operators, the exchange Hamiltonian (2) is
\be
\label{7}
\hat H_{ex} = -\; \frac{1}{2} \sum_{i\neq j} \left ( 
J_{ij} S_i^+ S_j^-  + I_{ij} S_i^z S_j^z \right ) \;   .
\ee

Writing down the Heisenberg equations of motion, we compliment them by the 
attenuation $\gamma_1$, caused by spin-lattice interactions and $\gamma_2$,
due to other spin interactions, e.g., dipole interactions that usually
are smaller than the exchange interactions. I that way, the equations of 
motion read as   
\be
\label{8}
 i\; \frac{dS_j^\pm}{dt} = \left [ S_j^\pm , \; \hat H \right ] -
i \gm_2 S_j^\pm \; , \qquad
 i\; \frac{dS_j^z}{dt} = \left [ S_j^z , \; \hat H \right ] -
i \gm_1 \left ( S_j^z - S_0 \right ) \; .
\ee
Accomplishing the related commutations, we get the equations for the 
transverse spin,
\be
\label{9}
\frac{dS_i^-}{dt} = i \sum_{j(\neq i)} \left ( I_{ij} S_i^- S_j^z - 
J_{ij} S_i^z S_j^- \right ) - i \om_0 S_i^- - 
i\mu_0 H S_i^z - \gm_2 S_i^- \; ,
\ee
and for the longitudinal spin
\be
\label{10}
 \frac{dS_i^z}{dt} = \frac{i}{2} \sum_{j(\neq i)} J_{ij}  \left (
S_i^+ S_j^- - S_i^- S_j^+ \right ) + \frac{ i}{2}\; \mu_0 H \left (
S_i^+ -  S_i^- \right ) - \gm_1 \left ( S_i^z - S_0 \right ) \;  .
\ee

\section{Stochastic mean-field approximation}

We shall be interested in the evolution of the averaged quantities 
characterizing the {\it transition function}
\be
\label{11}
u \equiv \frac{1}{NS} \sum_{j=1}^N \; \lgl S_j^- \rgl \;   ,
\ee
the {\it coherence intensity}
\be
\label{12}
 w \equiv \frac{1}{N(N-1)S^2} \sum_{i\neq j}^N \;
\lgl S_i^+ S_j^- \rgl \;  ,
\ee
and the {\it spin polarization}
\be
\label{13}
 z \equiv \frac{1}{NS} \sum_{j=1}^N \; \lgl S_j^z \rgl \;  .
\ee
 
Averaging equations (9) and (10), we invoke {\it stochastic mean-field 
approximation} \cite{Yukalov_19}, replacing spin pair correlators as
\be
\label{14}
 \lgl S_i^\al S_j^\bt \rgl \ra \lgl S_i^\al \rgl \lgl S_j^\bt \rgl 
+ \lgl S_i^\al \rgl \dlt S_j^\bt  + 
\lgl S_j^\bt \rgl \dlt S_i^\al \; ,
\ee
where $i \neq j$ and $\delta S_j^\alpha$ is treated as a stochastic variable, 
such that its stochastic average be zero:
\be
\label{15}
 \lgl \lgl \dlt S_j^\al \rgl \rgl = 0 \;  .
\ee
Then the averages $\langle S_j^\alpha \rangle$ become functions of the 
stochastic variables. 

Also, we introduce the composite stochastic variables
$$
\xi_0 \equiv \sum_{j(\neq i)} \left ( J_{ij} \dlt S_i^z - 
I_{ij} \dlt S_j^z \right ) \; , \qquad
\xi \equiv \sum_{j(\neq i)} \left ( I_{ij} \dlt S_i^- - 
J_{ij} \dlt S_j^- \right ) \; ,
$$
\be
\label{16}
 \vp \equiv \sum_{j(\neq i)} J_{ij} \left (  \dlt S_i^- - 
 \dlt S_j^- \right ) \; .
\ee
And we define the effective anisotropy
\be
\label{17}
 \Dlt J \equiv \frac{1}{N} \sum_{i\neq j} ( I_{ij} - J_{ij} ) \;  .
\ee
We assume that $\langle S_j^\alpha \rangle$ does not depend on the index 
enumerating spins and that the variables (16) also are not index-dependent. 

In this way, from equation (9), we obtain the equation for the transition 
function, 
\be
\label{18} 
 \frac{du}{dt} = - i ( \om_0 + \xi_0 - S \Dlt J z - i\gm_2 ) u -
i ( \mu_0 H - \xi ) z \;  ,
\ee
and for the coherence intensity,
\be
\label{19}
 \frac{dw}{dt} = - 2 \gm_2 w + 
i \left ( \mu_0 H - \xi^* \right ) z u - 
i \left ( \mu_0 H - \xi \right ) z u^* \; .
\ee
While equation (10) yields the equation for the spin polarization,
\be
\label{20}
 \frac{dz}{dt} = \frac{i}{2} \; u^*  \left ( \mu_0 H - \vp \right ) 
- \; \frac{i}{2}\; u \left ( \mu_0 H - \vp^* \right ) - 
\gm_1 ( z - \zeta ) \; ,
\ee
where $\zeta \equiv S_0/S$. The initial conditions
$$
u_0 = u(0) \; , \qquad w_0 = w(0) \; , \qquad
z_0 = z(0) \; , 
$$
compliment the above equations.

\section{Resonator feedback field}

The Kirchhoff equation (4) can be represented \cite{Yukalov_8,Yukalov_13} 
as the integral equation
\be
\label{21}
 H = - 4\pi \int_0^t G(t-t') \dot{m}_x(t') \; dt' \;  ,
\ee
with the transfer function
$$
 G(t) = \left [ \cos(\overline\om t) -\; 
\frac{\gm}{\overline\om} \; \sin(\overline\om t) \right ] \; e^{-\gm t} \; , 
\qquad  
\overline\om \equiv \sqrt{\om^2 - \gm^2} \; ,
$$
and where
\be
\label{22}
\dot{m}_x = \frac{\mu_0 NS}{2V_c} \; \frac{d}{dt} \; \left ( u^*
+ u \right ) \;   .
\ee
The longitudinal and transverse attenuations are assumed to be small,
as compared to the Zeeman frequency,
\be
\label{23}
 \frac{\gm_1}{\om_0} \ll 1 \; , \qquad \frac{\gm_2}{\om_0} \ll 1 \;  .
\ee
The coupling of the magnet with the resonant electric circuit leads to
the appearance of the {\it coupling attenuation}
\be
\label{24}
\gm_c \equiv \pi \mu_0^2 S \; \frac{N}{V_c} \;  .
\ee
The latter, together with the circuit attenuation, are small, as compared
to the resonator natural frequency,
\be
\label{25}
\frac{\gm}{\om} \ll 1 \; , \qquad   \frac{\gm_c}{\om} \ll 1 \; .
\ee
The stochastic variables (16), because of condition (15), can also be 
treated as effectively small.

To be efficient, the resonator has to be tuned to the Zeeman frequency, so 
that the resonance condition
\be
\label{26}
\frac{|\Dlt |}{\om} \ll 1 \qquad ( \Dlt \equiv \om - \om_0 ) 
\ee
be valid. Also, the magnetic anisotropy has to be small, with the anisotropy 
parameter 
\be
\label{27}
 A \equiv \frac{S \Dlt J}{\om_0} < 1 \;  .
\ee
Otherwise the sample magnetization would be frozen. 

In this way, the feedback equation (21) can be solved iteratively, taking
for the initial approximation $u \approx u_0 \exp(-i \omega_S t)$, with
\be
\label{28}
 \om_S \equiv \om_0 - S \Dlt J z = \om_0 ( 1 - Az ) \;  .
\ee
Then in the first iteration, we get
\be
\label{29}
\mu_0 H = i \left ( u \psi - u^* \psi^* \right ) \;   ,
\ee
with the coupling function
\be
\label{30}
\psi = \gm_c \om_S \left [ 
\frac{1 - \exp\{ -i(\om-\om_S)t-\gm t \} }{\gm+i(\om-\om_S)} + 
\frac{1 - \exp\{ -i(\om+\om_S)t-\gm t \} }{\gm-i(\om+\om_S)} 
\right ] \; .
\ee
In view of the above conditions, the first, resonant, term of the coupling
function prevails, so that 
\be
\label{31}
 \psi \cong \gm_c \; \om_S  \; 
\frac{1 - \exp\{ -i\Dlt_S t-\gm t\}}{\gm+i\Dlt_S} ,
\ee
with the effective dynamic detuning
\be
\label{32}
 \Dlt_S \equiv \om - \om_S = \Dlt + \om_0 A z \;  .
\ee

Defining the dimensionless coupling parameter
\be
\label{33}
 g \equiv \frac{\gm_c\om_0}{\gm\gm_2} \;  ,
\ee
the real and imaginary parts of the coupling function can be written as
\be
\label{34}
 {\rm Re}\psi = g \; \frac{\gm^2\gm_2}{\gm^2+\Dlt_S^2}\; ( 1 - Az)
\left \{ 1 - \left [ \cos(\Dlt_S t) -\; \frac{\Dlt_S}{\gm}\;
\sin(\Dlt_S t) \right ] e^{-\gm t} \right \}  
\ee
and, respectively,
\be
\label{35}
 {\rm Im}\psi = - g \; \frac{\gm\gm_2\Dlt_S}{\gm^2+\Dlt_S^2}\; ( 1 - Az)
\left \{ 1 - \left [ \cos(\Dlt_S t) + \frac{\gm}{\Dlt_S}\;
\sin(\Dlt_S t) \right ] e^{-\gm t} \right \}  \; .
\ee

Note that in the case of resonance, when $\Delta = 0$, and under weak 
anisotropy $A \ll 1$, we have $\Delta_S \ra 0$. Then the imaginary 
part (35) is close to zero, and the real part can be simplified to
$$
 {\rm Re}\psi \approx g \gm_2 ( 1 - A z ) \left ( 1 - 
e^{-\gm t} \right ) \;  .
$$
The latter form can be employed only when the anisotropy is not strong,
so that $A \ll 1$.

\section{Scale separation approach}

Substituting equality (28) into the evolution equations (18), (19), and (20),
we have the equations for the transverse function,
\be
\label{36}
 \frac{du}{dt} = - i ( \om_S + \xi_0 ) u - (\gm_2 - z \psi ) u -
z \psi^* u^* + i \xi z \;  ,
\ee
the coherence intensity,
\be
\label{37}
 \frac{dw}{dt} = - 2 ( \gm_2 - \al z ) w + 
i \left ( u^* \xi - \xi^* u \right ) z - 
z \left [ \psi u^2 + \psi^* (u^*)^2 \right ] \;  ,
\ee
and for the spin polarization,
\be
\label{38}
 \frac{dz}{dt} = - \al w -\; \frac{i}{2} 
\left ( u^*\vp - \vp^* u \right ) + 
\frac{1}{2} \left [ \psi u^2 + \psi^* ( u^*)^2 \right ] - 
\gm_1 ( z - \zeta ) \; .
\ee
Here the notation
\be
\label{39}
\al \equiv \frac{1}{2} \; \left ( \psi^* + \psi \right ) =
{\rm Re}\psi
\ee
is introduced. 

We solve the system of equations (36), (37), and (38) by resorting to the
scale separation approach
\cite{Yukalov_6,Yukalov_8,Yukalov_9,Yukalov_13,Yukalov_18,Yukalov_19}. 
The function $u$ is classified as fast, while $w$ and $z$, as slow. First,
we solve equation (36) for the fast variable, keeping the slow variables as 
quasi-integrals of motion and taking account of the resonance condition (26). 
This gives
$$
u = u_0 \exp \left \{ - ( i\om_S + \gm_2 -
z\psi ) t - i \int_0^t \xi_0(t') \; dt' \right \} \; +
$$
\be
\label{40}
+ \;
i z \int_0^t \xi(t') \exp \left \{ - ( i\om_S + \gm_2 - z\psi)(t - t') -
i \int_{t'}^t \xi(t'')\; dt'' \right \} \; dt' \; .
\ee

Stochastic variables, in view of condition (15), are zero-centered,
\be
\label{41}
 \lgl \lgl \xi_0(t) \rgl \rgl =  \lgl \lgl \xi(t) \rgl \rgl =
\lgl \lgl \vp(t) \rgl \rgl = 0 \; .
\ee
The stochastic variable $\xi_0$ is real, while $\xi$ and $\phi$ are
complex-valued. Therefore
\be
\label{42}
 \lgl \lgl \xi_0(t)\xi(t') \rgl \rgl =  
\lgl \lgl \xi_0(t)\vp(t') \rgl \rgl =
\lgl \lgl \xi(t)\vp(t') \rgl \rgl = 0 \; .
\ee
For the non-zero pair stochastic correlators, we set
\be
\label{43}
 \lgl \lgl \xi^*(t)\xi(t') \rgl \rgl = 2\gm_3 \dlt(t - t') \; ,
\qquad
\lgl \lgl \xi^*(t)\vp(t') \rgl \rgl = 2\gm_3 \dlt(t - t') \;  ,
\ee
with $\gamma_3$ playing the role of an attenuation caused by stochastic
fluctuations. 

Substituting expression (40) into equations (37) and (38) and averaging the
latter over fast temporal oscillations and stochastic variables, we come to
the equations for the guiding centers, describing the coherence intensity,
\be
\label{44}
 \frac{dw}{dt} = - 2 (\gm_2 - \al z ) w + 2\gm_3 z^2 
\ee
and the spin polarization,
\be
\label{45}
 \frac{dz}{dt} = - \al w - \gm_3 z - \gm_1 ( z - \zeta ) \; .
\ee
Recall that $\alpha$ is given by equations (34) and (39).

\section{Triggering spin waves}

Stochastic variables are the fluctuations that trigger spin motion. To 
clarify the nature of stochastic variables, let us show that these
correspond to spin waves describing spin fluctuations around the average
local spin polarization $z = z(t)$. It is worth recalling that spin waves
can be well defined for nonequilibrium systems \cite{Ruckriegel_20,Birman_21}. 

Spin fluctuations are defined as small oscillations around the average spin,
which can be represented as
\be
\label{46}
 S_j^\al = \lgl S_j^\al \rgl + \dlt S_j^\al \; .
\ee
Substituting this into the equations of motion (9) and (10), we omit, for 
simplicity, the attenuations, keeping in mind the initial stage of the 
process, when time is yet much shorter than the relaxation times. 
Separating the zero-order equations, we get
\be
\label{47}
 \frac{d}{dt} \; \lgl S_j^- \rgl = - i \om_S \lgl S_j^- \rgl \; ,
\qquad  \frac{d}{dt} \; \lgl S_j^z \rgl = 0 \; .
\ee
And to first order, we have
$$
\frac{d}{dt} \; \dlt S_j^-  = - i \left ( \om_0 \dlt S_j^- +
\lgl S_j^- \rgl \xi_0 - \lgl S_j^z \rgl \xi \right ) \; ,
$$
\be
\label{48}  
 \frac{d}{dt} \; \dlt S_j^z  = \frac{i}{2} \; \left ( \lgl S_j^- \rgl
\vp^* - \lgl S_j^+ \rgl \vp  \right ) \;  .
\ee
Keeping in mind that $z$ is a slow variable, from equations (47), we find
\be
\label{49}
 \lgl S_j^- \rgl = u_0 S e^{-i\om_S t} \; , \qquad
 \lgl S_j^z \rgl = z S \; .
\ee

In order to stress the role of fluctuations, let us set $u_0 = 0$. Then
\be
\label{50}
\dlt S_j^- = S_j^- \; , \qquad \dlt S_j^z = 0 \qquad 
( u_0 = 0 ) \;  ,
\ee
where we take into account condition (15). The first of equations (48) 
yields 
\be
\label{51}
 \frac{d}{dt} \;  S_j^- = - i\om_0 S_j^- + i z S \xi \;  .
\ee

Let us employ the Fourier transformation for the ladder spin operators,
$$
 S_j^- = \sum_k S_k^- e^{i\bk\cdot\br_j} \; , \qquad
 S_k^- = \frac{1}{N} \sum_j S_j^- e^{-i\bk\cdot\br_j} \;  ,
$$
and for the exchange interactions,
$$
 J_{ij} = \frac{1}{N} \sum_k J_k e^{i\bk\cdot\br_{ij}} \;  ,
\qquad
J_k =  \sum_{j(\neq i)} J_{ij} e^{-i\bk\cdot\br_{ij}} \;   ,
\qquad ( \br_{ij} \equiv \br_i - \br_j)
$$
with the similar transformation for $I_{ij}$. For the stochastic variable 
$\xi$, we get
$$
\xi = \sum_k ( I_0 - J_k ) S_k^- e^{i\bk \cdot\br_j} \;   .
$$
Then equation (51) reduces to
\be
\label{52}
 \frac{d}{dt} \; S_k^- = - i\om_k S_k^- \;  ,
\ee
with the spin-wave spectrum 
\be
\label{53}
  \om_k = \om_0 + z S ( J_k - I_0 ) \; .
\ee
For long waves, when $k \ra 0$, we obtain
\be
\label{54}
 \om_k \simeq \om_S - \; \frac{zS}{2} 
\sum_{j(\neq i)} J_{ij} (\bk \cdot\br_{ij})^2 \;  .
\ee
This demonstrates that stochastic fluctuations are nothing but spin waves. 

For what follows, it is useful to keep in mind the restriction
\be
\label{55}
 w + z^2 = \left | \; \frac{1}{NS} 
\sum_{j=1}^N \; \lgl \bS_j \rgl \; \right |^2 \leq 1 \;  ,
\ee
which is necessary to take into account when setting initial conditions for
equations (44) and (45).

\section{Qualitative classification of regimes}

Suppose that the initial state of the magnet is nonequilibrium, such that 
the external magnetic field is directed along the $z$-axis, together with
spins. Spin waves trigger spin motion, forcing spins to move, which 
generates the resonator feedback field acting back on spins and 
collectivizing their motion. The spins tend to reach an equilibrium state,
reversing from positive to negative values. The overall dynamics can be
classified onto several qualitatively different stages. Below, we give this
qualitative classification, setting for simplicity zero detuning $\Delta = 0$
and considering the limit of negligibly weak anisotropy $A \ra 0$. Thus, we 
set
\be
\label{56}
 \Dlt \ra 0 \; , \qquad A \ra 0 \; , \qquad \Dlt_S \ra 0 \;  .
\ee
Then the coupling function (39) reduces to the simple form 
\be
\label{57}
 \al \ra g\gm_2 \left ( 1 - e^{-\gm t} \right ) \;  .
\ee

\subsection{Chaotic stage}

At the beginning, before coherence in spin motion sets up at time $t_{coh}$, 
the process is yet chaotic. At short times in the interval
\be
\label{58}
 0 < t < t_{coh} \;  ,
\ee
the coupling function is yet very small, such that
\be
\label{59}
  \al \ll \gm_2 \; , \qquad \al \ll \gm_3 \; .
\ee
Then the linear in time solutions of equations (44) and (45) are
\be
\label{60}
 w \simeq w_0 + 2 \left ( \gm_3 z_0^2 - \gm_2 w_0 \right ) t \; ,
\qquad
z \simeq z_0 - \left [ ( z_0 - \zeta ) \gm_1 + 
\gm_3 z_0 \right ] t \;  .
\ee

This regime lasts till the time, when the coupling function grows, so that 
\be
\label{61}
 \al z = \gm_2 \qquad ( t = t_{coh} ) \;  ,
\ee
which defines the {\it coherence time}  
\be
\label{62}
t_{coh} = \frac{1}{\gm} \; \ln \; \frac{gz_0}{gz_0 -1 } \; .
\ee
At large coupling parameter, this gives
\be
\label{63}
 t_{coh} =  \frac{\gm_2}{\gm_c \om_0 z_0} \qquad ( gz_0 \gg 1 ) \;  .
\ee

\subsection{Coherent stage}

After the coherence time, spins become well correlated by means of the 
resonator feedback field. In the time interval
\be
\label{64}
 t_{coh} < t < T_2 \;  ,  
\ee
where $T_2 \equiv 1/ \gamma_2$, the coupling function is large,
\be
\label{65}
 \al \gg \gm_1 \; , \qquad \al \gg \gm_3 \;  ,
\ee
reaching the value
\be
\label{66}
 \al \simeq g \gm_2  \qquad (\gm_2 < \gm ) \;  .
\ee

The spin-lattice attenuation is usually small, such that $\gm_1\ll \gm_2$
and $\gm_1\ll \gm_3$. Then equations (44) and (45) read as
\be
\label{67}
 \frac{dw}{dt} = - 2\gm_2 ( 1  - gz) w \; , \qquad
\frac{dz}{dt} = - g\gm_2 w \;  .
\ee
The exact solutions of these equations give the coherence intensity
\be
\label{68}
w = \left ( \frac{\gm_p}{g\gm_2} \right )^2 {\rm sech}^2 \left (
\frac{t-t_0}{\tau_p} \right ) 
\ee
and the spin polarization
\be
\label{69}
 z = -\; \frac{\gm_p}{g\gm_2}  \tanh \left (
\frac{t-t_0}{\tau_p} \right ) + \frac{1}{g} \;  .
\ee
Here $t_0$ and $\tau_p$ are the integration constants that are defined by 
the values $w(t_{coh}) \equiv w_{coh}$ and $z(t_{coh}) \equiv z_{coh}$. 
This gives the {\it delay time}
\be
\label{70}
t_0 = t_{coh} + \frac{\tau_p}{2} \; 
\ln \left | \; \frac{\gm_p+\gm_g}{\gm_p-\gm_g} \; \right |
\ee
and the {\it pulse time}
\be
\label{71}
  \tau_p = \frac{1}{\gm_p} \; ,
\ee
where
\be
\label{72}
 2\gm_p^2 = \gm_g^2 \left [ 1 + 
\sqrt{ 1 + 4 \left ( \frac{g\gm_2}{\gm_g} \right )^2 w_{coh} } 
\right ]\; , \qquad 
\gm_p = \gm_2 ( g z_{coh} -1 ) \;  .
\ee
Under a large coupling parameter and small coherence intensity, at the 
coherence time $t_{coh}$, we get
$$
\gm_g \simeq g\gm_2 z_{coh} \qquad ( gz_{coh} \gg 1 ) \; ,
$$
\be
\label{73}
 \gm_p \simeq g\gm_2\; \sqrt{ z^2_{coh} + w_{coh}} \qquad 
( w_{coh} \ll z_{coh} ) \;  .
\ee
The maximum of the coherence intensity occurs at the time $t_0$, being
\be
\label{74}
 w(t_0) = w_{coh} + \left ( z_{coh} -\; \frac{1}{g} \right )^2 \; .
\ee
At this moment of time, the spin polarization is
\be
\label{75}
 z(t_0) = \frac{1}{g} \; .
\ee

\subsection{Relaxation stage}

After the transverse decoherence time $T_2$, there is the relaxation stage
in the interval
\be
\label{76}
T_2 < t < T_1 \; ,
\ee
where $T_1 \equiv 1/ \gamma_1$. For $t \gg t_0$, the coherence intensity 
relaxes as
\be
\label{77}
 w \simeq 4w(t_0) \exp\left ( - \; \frac{2t}{\tau_p} \right ) \; ,
\ee
and the spin polarization tends to
\be
\label{78} 
 z \simeq - z_{coh} + \frac{2}{g} + 
2 \left ( z_{coh} -\; \frac{1}{g} \right ) 
\exp \left ( - \; \frac{2t}{\tau_p} \right ) \;  .
\ee

\subsection{Quasi-stationary stage}

In the limit of very long times, when
\be
\label{79}
 T_1 < t < \infty \;  ,
\ee
the solutions exhibit small oscillations around the fixed points $w^*$ 
and $z^*$ given by the equations
\be
\label{80}
 \gm_2 ( 1 - gz^* ) w^* - \gm_3 ( z^*)^2 = 0 \; , \qquad
g\gm_2 w^* + \gm_3 z^* + \gm_1 ( z^* - \zeta ) = 0 \;  .
\ee

\subsection{Punctuated superradiance}

The coherent stage of radiation corresponds to the regime of superradiance,
which can be pure, if $w_0 = 0$, or triggered, if $w_0 > 0$. It is also 
possible to realize the regime of punctuated superradiance \cite{Yukalov_18},
when in the process of spin dynamics either the external magnetic field is
reversed or the magnetic sample is rotated, so that to reproduce the initial
nonequilibrium conditions. In this case, radiation exhibits a series of 
coherent pulses, with the temporal intervals that can be regulated.

\section{Numerical investigation of dynamics}

We solve equations (44) and (45) numerically, with the coupling function 
defined in equations (34) and (39). The case of exact resonance is assumed, 
with $\omega = \omega_0$. Then $\Delta = 0$ and $\Delta_S = \omega A z$. 
The chosen initial conditions correspond to a purely self-organized 
process, in the absence of imposed initial coherence, so that $w_0 = 0$,
and when the initial polarization $z_0 = 1$ defines a strongly 
nonequilibrium state, since the equilibrium polarization of a single
spin, under the given setup, corresponds to $ \zeta = -1$. 

In the presence of a resonator feedback field, spin reversal happens much 
faster than the homogeneous transverse relaxation time $T_2 \equiv 1/\gamma_2$.
We measure all attenuations and frequencies in units of $\gamma_2$. The 
spin-lattice attenuation $\gamma_1$ is usually much smaller than $\gamma_2$.
Taking this into account, we set $\gamma_1 = 0.001$. The dynamic attenuation
$\gamma_3$ is assumed to be of order of $\gamma_2$, which, in units of the 
latter, implies $\gamma_3 = 1$. Time is measured in units of $T_2$.      
 
First, we consider the more general form corresponding to expression (34),
which reads as
$$
 \al = g \; \frac{\gm^2 \gm_2( 1 - Az)}{\gm^2 + (\om A z)^2} \;
\left \{ 1 - \left [ \cos(\om A z t) - \; \frac{\om}{\gm} \;
Az \sin( \om A z t) \right ] e^{-\gm t} \right \} \; .
$$
The temporal behaviour of the coherence intensity and spin polarization,
for different system parameters, is shown in Figures 1 to 8.

Figures 1 and 2 show the role of magnetic anisotropy for different 
resonator frequencies. Increasing the anisotropy, generally, shifts the 
delay time, widens the coherence pulse, and decreases the coherence peak 
maximum, and the reversed spin polarization. 

Figures 3 and 4 demonstrate the role of the resonator attenuation. The 
larger $\gamma$, the higher the coherence peak maximum and the shorter 
the delay time. Spin reversal is better pronounced for larger attenuations.    

Figures 5 and 6 illustrate the role of the coupling strength between 
the magnet and resonator. The larger coupling parameter leads to a 
shorter delay time and higher coherence maximum. The value of $\omega$
does not influence much the behavior, when the anisotropy is weak, 
$A \ll 1$. Stronger coupling leads to a better spin reversal.  

Figures 7 and 8 show the role of anisotropy for a large value of the
coupling parameter. The stronger anisotropy increases the delay time 
and diminishes the coherence peak maximum. Spin reversal is more effective
for a weaker anisotropy.

The general form of the above coupling function looks a bit cumbersome, 
because of which we also check its approximate form, discussed in 
Section 4, which is given by the expression
$$
\al \approx g\gm_2 ( 1 - Az) \left ( 1 - e^{-\gm t} \right ) \; .
$$
It turns out that this approximate form is applicable under weak anisotropy,
with the anisotropy parameters $A \ll 1$, but becomes invalid for larger
anisotropies.

\section{Spatial distribution of radiation}

The spatial distribution of radiation for atomic systems can be found 
in references \cite{Rehler_22,Allen_23}. In the case of spin systems, the 
calculational procedure is similar, with the difference that the radiation 
is produced by moving magnetic moments. The other slight difference is 
connected with the chosen geometry. For atomic systems, one usually 
considers a cylindric sample, with the axis along the $z$ axis. In the 
setup, related to a magnetic system, as we study here, the magnetic sample 
is inserted into a coil, with the axis along the $x$ axis, while an 
external magnetic field defines the $z$ axis, so that the sample axis is 
orthogonal to the $z$ axis. 

The operator of magnetic moment, related to a $j$-th spin, can be written 
in the form
\be
\label{81} 
\mu_0 \bS_j = \vec{\mu} S_j^+ + \vec{\mu}^* S_j^- +
\vec{\mu}_0 S_j^z \;  ,
\ee
where
\be
\label{82}
 \vec{\mu} = \frac{\mu_0}{2} \;  ( \bbe_x - i \bbe_y)  \; , 
\qquad \vec{\mu} = \mu_0 \bbe_z \; .
\ee
The radiation intensity in the direction of the vector
\be
\label{83}
\bbn \equiv \frac{\br}{|\br|} = \frac{\br}{r}
\ee
can be represented \cite{Yukalov_19} as
\be
\label{84}
 I(\bbn,t) = 2 \om_0 \gm_0 \sum_{ij} \vp_{ij}(\bbn)
\lgl S_i^+(t) S_j^-(t) \rgl \;  ,
\ee
where $\gamma_0$ is a natural width,
\be
\label{85}
 \gm_0 \equiv \frac{2}{3} \; |\vec{\mu}|^2 k_0^3 = 
\frac{1}{3} \; \mu_0^2 k_0^3  \qquad 
\left ( k_0 \equiv \frac{\om_0}{c} \right ) \; ,
\ee
and the system form-factor is
\be
\label{86}
\vp_{ij}(\bbn) = \frac{3}{8\pi} \; 
| \bbn \times \bbe_\mu |^2 \exp ( i k_0 \bbn \cdot \br_{ij} ) \;   .
\ee
Here we introduce the unit vector
\be
\label{87}
\bbe_\mu \equiv \frac{\vec{\mu}}{|\vec{\mu}|} = 
\frac{1}{\sqrt{2}} \; ( \bbe_x - i \bbe_y )
\ee
and take into account that
$$
 |\vec{\mu}| = \frac{\mu_0^2}{2} \; , \qquad 
|\vec{\mu}_0|^2 = \mu_0^2 \;  .
$$
Denoting by $\vartheta$ the angle between $\bf{n}$ and $\bf{e}_z$, 
we have 
$$
| \bbn \times \bbe_\mu |^2 = 1 - \; \frac{1}{2} \; \sin^2\vartheta =
\frac{1}{2} \; \left ( 1 + \cos^2\vartheta \right ) \; .
$$
The form factor (86) becomes
\be
\label{88}
 \vp_{ij}(\bbn) = \frac{3}{16\pi} \; \left ( 1 + \cos^2\vartheta \right )
\exp ( i k_0 \bbn \cdot \br_{ij} ) \;  .
\ee

The radiation intensity (84) can be separated into two terms, 
\be
\label{89}
I(\bbn,t) = I_{inc}(\bbn,t) + I_{coh}(\bbn,t) \;   ,
\ee
the incoherent radiation intensity
\be
\label{90}
 I_{inc}(\bbn,t) = 2\om_0 \gm_0 
\sum_j \vp(\bbn) \lgl S_j^+(t) S_j^-(t) \rgl \;  ,
\ee
and the coherent radiation intensity
\be
\label{91} 
  I_{coh}(\bbn,t) =  2\om_0 \gm_0 
\sum_{i\neq j} \vp_{ij}(\bbn) \lgl S_i^+(t) S_j^-(t) \rgl \;  ,
\ee
where
\be
\label{92}
 \vp(\bbn) \equiv \vp_{jj}(\bbn) = \frac{3}{16\pi} \; \left ( 1 +
\cos^2\vartheta \right ) \;  .
\ee

Taking into account the identity
$$
S_j^+ S_j^- = S ( S + 1) - \left ( S_j^z \right )^2 + S_j^z   
$$
and the approximate equality
$$
 \lgl \left ( S_j^z \right )^2 \rgl \approx S^2 \;  ,
$$
which is exact for spin $S = 1/2$, as well as for large $S \ra \infty$,
makes it possible to rewrite the incoherent radiation intensity as
\be
\label{93}
I_{inc}(\bbn,t) = 2\om_0 \gm_0 
\sum_j \vp(\bbn) \left ( S + \lgl S_j^z \rgl \right ) \; .
\ee

Introducing the local functions
$$
u_j(t) \equiv \frac{1}{S} \;  \lgl S_j^-(t) \rgl\; ,
\qquad
w_j(t) \equiv \frac{1}{S^2} \; | \; \lgl S_j^-(t) \rgl \; |^2 \; ,
$$
\be
\label{94}
 z_j(t) \equiv \frac{1}{S} \;  \lgl S_j^z(t) \rgl \; , 
\ee
and using the semiclassical approximation, we come to the incoherent 
radiation intensity
\be
\label{95}
I_{inc}(\bbn,t) =  2\om_0 \gm_0 S
\sum_j \vp_{ij}(\bbn) [ 1 + z_j(t) ] \; ,
\ee
and the coherent radiation intensity
\be
\label{96}
 I_{coh}(\bbn,t) =  2\om_0 \gm_0 S^2
\sum_{i\neq j} \vp_{ij}(\bbn) u_i^*(t) u_j(t) \;  .
\ee

Assuming that the radiation wave length is comparable or larger than the
linear sample sizes, we can resort to the uniform approximation, passing to 
functions (11), (12), and (13). In this procedure, we consider the sum
\be
\label{97}   
 \sum_{i\neq j} \vp_{ij}(\bbn) = \vp(\bbn) N^2 \left [ F(k_0\bbn)
- \; \frac{1}{N^2} \right ] \;  ,
\ee
in which 
\be
\label{98}
 F(k_0\bbn) \equiv \left | \; \frac{1}{N} 
\sum_{j=1}^N e^{ik_0\bbn \cdot\br_j} \; \right |^2 \;   .
\ee
Then we find the incoherent radiation intensity
\be
\label{99}
I_{inc}(\bbn,t) = 2\om_0\gm_0 N S \vp(\bbn) [ 1 + z(t) ]
\ee
and the coherent radiation intensity
\be
\label{100}
 I_{coh}(\bbn,t) = 
2\om_0\gm_0 N^2 S^2 \vp(\bbn) F(k_0\bbn) w(t) \; .
\ee

The total radiation intensity, integrated over the spherical angles, is
\be
\label{101}
 I(t) \equiv \int I(\bbn,t) \; d\Om(\bbn) =
I_{inc}(t) + I_{coh}(t) \;  ,
\ee
containing the incoherent part
\be
\label{102}
I_{inc}(t) = 2\om_0\gm_0 SN [ 1 + z(t) ] 
\ee
and the coherent part
\be
\label{103}
 I_{coh}(t) = 2\om_0\gm_0 S^2 N^2 \vp_0 w(t) \; ,
\ee
with the shape factor
\be
\label{104}
 \vp_0 \equiv \int \vp(\bbn) F(k_0\bbn) \; d\Om(\bbn) \; .
\ee

When spins are uniformly distributed in the sample, function (98) can be
represented as an integral over the sample volume: 
\be
\label{105}
 F(k_0\bbn) = \left | \; \frac{1}{V} \int_V e^{ik_0\bbn \cdot\br} \;
d\br \; \right |^2 \;  .
\ee
The direction vector, in terms of spherical coordinates, is
\be
\label{106}
 \bbn = \{ \sin\vartheta \cos\vp , \;  \sin\vartheta \sin\vp , \; 
\cos\vartheta \} \; .
\ee
Integrating over the sample, we meet the integral
$$
\int_0^a \sin\left ( c\sqrt{a^2-x^2} \right ) \cos(bx) \; dx =
\frac{\pi a c}{2\sqrt{b^2+c^2}} \; J_1 
\left ( a \sqrt{b^2+c^2} \right )
$$
expressed through the Bessel function $J_1$ of the first kind. Finally,
we obtain
\be
\label{107}
  F(k_0\bbn) = 
\frac{16\sin^2\left ( \frac{k_0L}{2}\sin\vartheta\cos\vp\right )}
{k_0^2L^2\sin^2\vartheta\cos^2\vp} \;
\frac{J_1^2(k_0R\sqrt{\sin^2\vartheta\sin^2\vp+\cos^2\vartheta})}
{k_0^2R^2(\sin^2\vartheta\sin^2\vp+\cos^2\vartheta)} \; .
\ee
Recall that $k_0 \equiv \omega_0/c = 2 \pi/ \lambda$. 

Since the magnetic sample is inserted into a coil aligned along the axis 
$x$, the radiation in the $z$ direction is absorbed by the coil. Radiation 
can be emitted only in the $x$ direction, when
\be
\label{108}
 F(k_0\bbn) = \frac{4}{k_0^2L^2} \; 
\sin^2 \left ( \frac{k_0L}{2} \right ) \qquad 
\left ( \vartheta = \frac{\pi}{2} , \; \vp=0 \right ) \;  .
\ee
Then $\varphi (\bbn) = 3/(16 \pi)$, which is only twice smaller than its
value in the $z$ direction. 

If the sample is a very narrow cylinder, or represents a linear chain of spins
aligned along the axis $x$, then expression (107) leads to
\be
\label{109}
 F(k_0\bbn) = 
\frac{4\sin^2\left(\frac{k_0L}{2}\sin\vartheta\cos\vp\right)}
{k_0^2L^2\sin^2\vartheta\cos^2\vp} \qquad ( R \ra 0 ) \;  .
\ee

Magnetic dipole radiation usually corresponds to long waves, such that the
wavelength $\lambda$ is longer that the sample linear sizes. To consider 
this limit, we notice that the Bessel function $J_{\nu}(x)$, at small 
$x \ll 1$, has the asymptotic property
$$
J_\nu(x) \simeq \frac{1}{\Gm(\nu+1)} \left ( \frac{x}{2}\right )^\nu
- \; \frac{1}{\Gm(\nu+2)} \; \left ( \frac{x}{2} \right )^{\nu+2} \;  ,
$$
because of which
$$
 J_1(x) \simeq \frac{x}{2} \qquad ( x \ra 0 ) \;  .
$$
Hence, in the long-wave limit,
\be
\label{110}
F(k_0\bbn) \simeq 1 \qquad \left ( \frac{2\pi R}{\lbd} \ll 1 , \;\;
\frac{\pi L}{\lbd} \ll 1 \right ) \; .
\ee
     
In this way, although a part of radiation is absorbed by the coil, 
nevertheless, an essential part of radiation can be emitted through the 
sides of the sample along the coil axis $x$.

\section{Discussion}

We have considered a large class of magnetic materials, composed of 
particles interacting through exchange interactions. Such materials are 
well represented by an anisotropic Heisenberg Hamiltonian. By coupling 
a magnetic sample to a resonant electric circuit it is possible to 
efficiently regulate spin dynamics. The present investigation complements
the earlier studies of spin dynamics in different materials, accomplished 
for polarized nuclei, magnetic nanomolecules, and magnetic nanoclusters. 
The systems composed of these particles possess several common properties.
Such systems, except polarized nuclei, exhibit magnetic anisotropy. For 
example, the spins of magnetic nanomolecules are frozen below the blocking
temperature $T_B \sim 1 - 10$ K, with the frozen magnetization protected
by an anisotropy barrier of energy $E_A \sim 10 - 100$ K. For magnetic 
nanoclusters, the blocking temperature is $T_B \sim 10 - 100$ K. Polarized
nuclei can be represented by polarized protons in hydrogenated materials,
such as propanediol, butanol, and ammonia. Although these materials do not 
enjoy magnetic anisotropy, however their spins, at low temperature, can 
remain polarized for extremely long time. The common feature of all above 
materials is that their spin interactions are described by dipolar forces.          

However, there exists a very wide class of materials made of particles 
with magnetic exchange interactions. Also, recently there have appeared 
a new type of magnetic nanomaterials characterized by exchange interactions,
such as magnetic graphene, where magnetic properties are induced by defects 
\cite{Yaziev_2,Katsnelson_3,Enoki_4}. 

Graphene is a two-dimensional carbon material intermediate between an 
insulator and a metal \cite{Goerbig_24,Wehling_25}. Carbon-carbon spacing 
is $a \approx 1.42$ \AA, and its surface density is 
$\rho \approx 3.9 \times 10^{15}$ cm$^{-2}$.

Magnetic graphene can be well described by an anisotropic Heisenberg model
\cite{Yaziev_2,Katsnelson_3,Enoki_4}. For defects on a zigzag edge, one has
an exchange interaction potential $J \sim 0.1$ eV, or $J \sim 10^{-13}$ erg. 
This gives $J/\hbar \sim 10^{14}$ s$^{-1}$ and $J/k_B \sim 10^3$ K. Dipolar 
interactions are much weaker, with $\mu_B^2/a^3 \sim 10^{-17}$ erg, or
$\mu_B^2/(\hbar a^3) \sim 10^{10}$ s$^{-1}$. In temperature units, this 
gives $\mu_B^2/(k_B a^3) \sim 0.1$ K. Magnetic anisotropy is not strong, 
with $\Delta J/ J\sim 10^{-4}$, hence, $\Delta J/ \hbar \sim 10^{10}$ s$^{-1}$. 
Thus, $\gamma_2 \sim \mu_B^2/(\hbar a^3) \sim 10^{10}$ s$^{-1}$. If 
$\gamma_3 \sim \Delta J/ \hbar$, then $\gamma_3 \sim 10^{10}$ s$^{-1}$,
that is, of the same order as $\gamma_2$. In the case of an external magnetic
field $B_0 = 1$ T, the Zeeman frequency is 
$\omega_0 = |\mu_B B_0|/ \hbar \sim 10^{11}$ s$^{-1}$. Therefore the radiation
wavelength is $\lambda \sim 10$ cm. Since there are many ways of generating 
defects in graphene, the system parameters can be varied. 

It is important to stress that a self-organized spin dynamics, when there 
is no initial coherence imposed onto the sample, cannot be correctly described 
by phenomenological equations, like Landau-Lifshitz, Gilbert, or Bloch 
equations. This is why, we use a microscopic approach with a realistic 
Hamiltonian. Such an approach makes it possible to take into account quantum
fluctuations, crucially important at the beginning of spin relaxation. Also, 
the use of a microscopic model has an advantage of containing well defined
parameters associated with the considered Hamiltonian.

When the system is prepared in a nonequilibrium state, and no initial 
coherence is imposed onto the sample, spin wave fluctuations serve as a 
triggering mechanism starting spin motion. It is possible to show that 
spin wave fluctuations is the sole triggering mechanism, while thermal 
Nyquist noise of the coil cannot play the role of a trigger. 

In the process of motion, spins are collectivized by means of the resonator
feedback field. Superradiance in spin systems cannot be caused by photon 
exchange. That is, spin superradiance is due to the Purcell effect and is 
impossible without a resonator. This is contrary to atomic systems, where 
superradiance develops as a Dicke effect.      
        
The microscopic equations of motion are investigated by employing stochastic
mean-field approximation and scale separation approach. The equations for 
guiding centers are solved numerically. Depending on initial conditions and 
system parameters, there can exist different regimes of spin dynamics, slow
free relaxation during the time $T_1$, free induction in time $T_2$, weak
superradiance in time slightly shorter tan $T_2$, pure superradiance and
triggered superradiance, when the reversal time of magnetization can be made 
much shorter than $T_2$, of order $10^{-11}$ s or $10^{-12}$ s. The regime
of punctuated superradiance can be realized, producing a sequence of coherent
pulses. 
      
We have analyzed the spatial distribution of radiation produced by moving 
magnetic moments. Although a part of radiation is absorbed by the coil
surrounding the sample, anyway, radiation can be emitted through the sides
of the sample, where there is no coil.

The studied effects can be used in a variety of problems in spintronics and
in quantum information processing.

\vskip 2mm

{\bf Acknowledgement}. Financial support from RFBR (grant $\#$ 14-02-00723)
is appreciated.

\newpage
\begin{center}
{\Large{\bf Figure Captions }}

\end{center}

\vskip 2cm

{\bf Figure 1}. Role of magnetic anisotropy in spin dynamics for the 
parameters $\gamma = 1$, $g = 10$, and $\omega = 10$, under different 
anisotropy parameters, $A = 0.1$ (solid line), $A = 0.5$ (dashed line), 
and $A = 1$ (dashed-dotted line): (a) Coherence intensity; 
(b) Spin polarization. 

\vskip 1cm
{\bf Figure 2}. Role of magnetic anisotropy in spin dynamics for the 
parameters $\gamma = 1$, $g = 10$, but $\omega = 100$, under different 
anisotropy parameters, $A = 0.1$ (solid line), $A = 0.5$ (dashed line), 
and $A = 1$ (dashed-dotted line): (a) Coherence intensity; 
(b) Spin polarization. 

\vskip 1cm
{\bf Figure 3}. Role of resonator attenuation for the parameters 
$g = 10$, $A = 0.1$, and $\omega = 10$, under different attenuation 
parameters, $\gamma = 1$ (solid line), $\gamma = 10$ (dashed line), 
and $\gamma = 100$ (dashed-dotted line): (a) Coherence intensity;
(b) Spin polarization.

\vskip 1cm
{\bf Figure 4}. Role of resonator attenuation for the parameters 
$g = 10$, $A = 0.1$, but $\omega = 100$, under different attenuation 
parameters, $\gamma = 1$ (solid line), $\gamma = 10$ (dashed line), 
and $\gamma = 100$ (dashed-dotted line): (a) Coherence intensity;
(b) Spin polarization.

\vskip 1cm
{\bf Figure 5}. Role of magnet-resonator coupling for the parameters 
$\gamma = 10$, $A = 0.1$, and $\omega = 10$, under different coupling 
parameters $g = 10$ (solid line) and $g = 100$ (dashed line):
(a) Coherence intensity; (b) Spin polarization.

\vskip 1cm
{\bf Figure 6}. Role of magnet-resonator coupling for the parameters 
$\gamma = 10$, $A = 0.1$, but $\omega = 100$, under different coupling 
parameters, $g = 10$ (solid line) and $g = 100$ (dashed line):
(a) Coherence intensity; (b) Spin polarization.

\vskip 1cm
{\bf Figure 7}. Role of magnetic anisotropy for the parameters $\gamma = 10$,
$g = 100$, and $\omega = 10$, under different anisotropy parameters,
$A = 0$ (solid line), $A = 0.1$ (dashed line), $A = 0.5$ 
(dashed-dotted line), and $A = 1$ (dashed line with dots): (a) Coherence 
intensity; (b) Spin polarization.

\vskip 1cm
{\bf Figure 8}. Role of magnetic anisotropy for the parameters $\gamma = 10$,
$g = 100$, but $\omega = 100$, under different anisotropy parameters,
$A = 0$ (solid line), $A = 0.1$ ( dashed line), $A = 0.5$ 
(dashed-dotted line), and $A = 1$ (dotted line): (a) Coherence 
intensity; (b) Spin polarization.

\newpage

%Figure 1
\begin{figure}[ht]
\vspace{9pt}
\centerline{
\hbox{ \includegraphics[width=7.5cm]{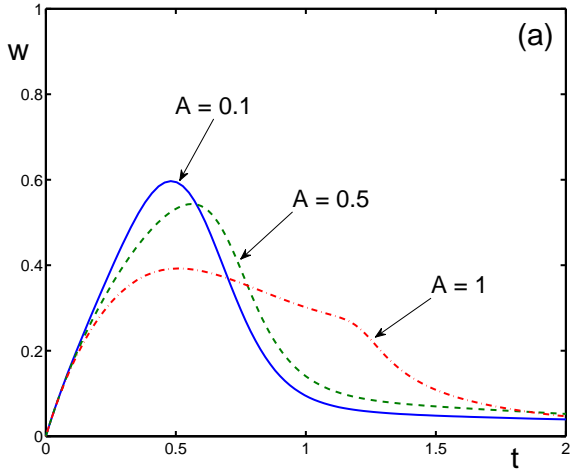} \hspace{2cm}
\includegraphics[width=7.5cm]{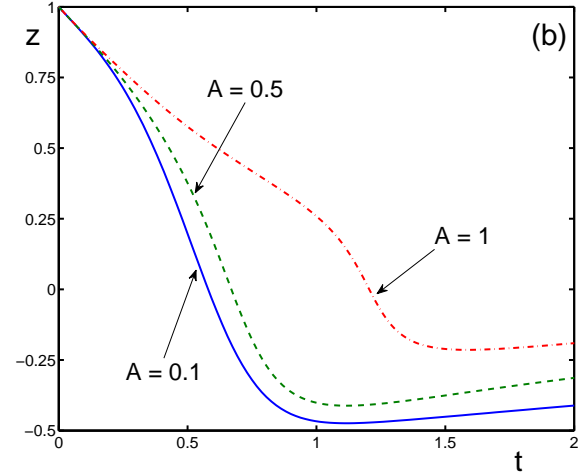} } }
\caption{Role of magnetic anisotropy in spin dynamics for the
parameters $\gamma = 1$, $g = 10$, and $\omega = 10$, under different
anisotropy parameters, $A = 0.1$ (solid line), $A = 0.5$ (dashed line),
and $A = 1$ (dashed-dotted line): (a) Coherence intensity;
(b) Spin polarization.
}
\label{fig:Fig.1}
\end{figure}

\vskip 3cm

%Figure 2
\begin{figure}[ht]
\vspace{9pt}
\centerline{
\hbox{ \includegraphics[width=7.5cm]{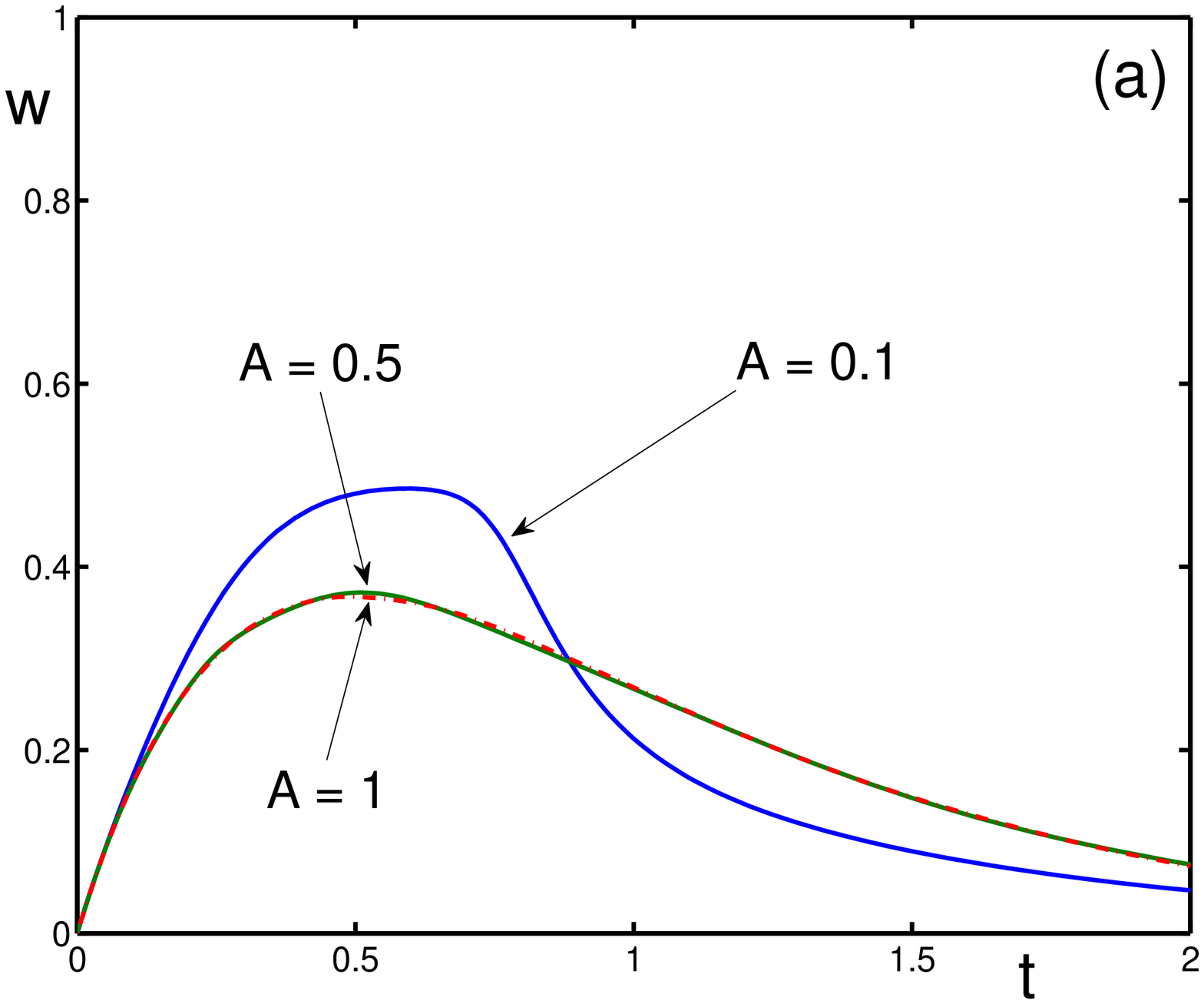} \hspace{2cm}
\includegraphics[width=7.5cm]{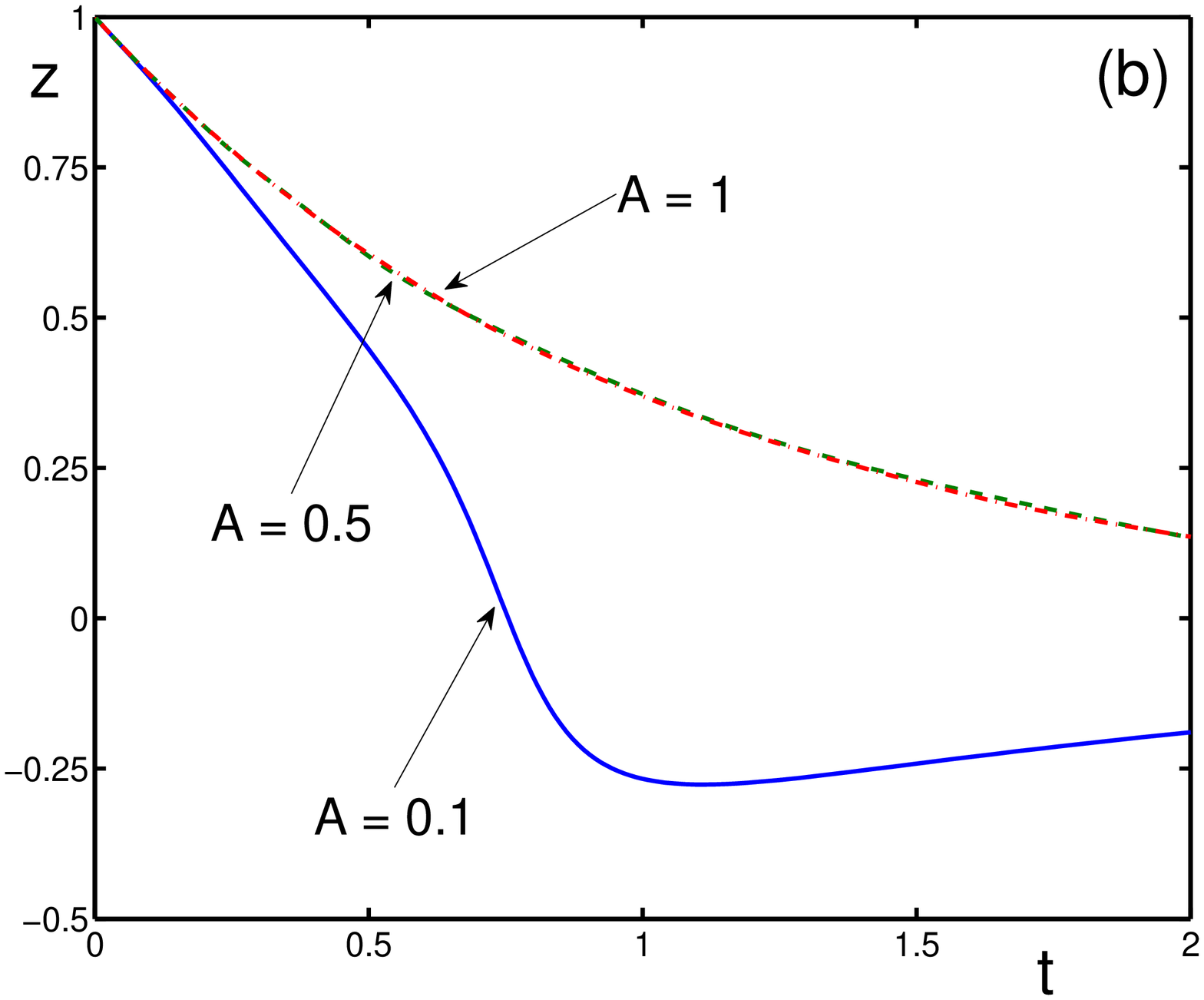} } }
\caption{Role of magnetic anisotropy in spin dynamics for the
parameters $\gamma = 1$, $g = 10$, but $\omega = 100$, under different
anisotropy parameters, $A = 0.1$ (solid line), $A = 0.5$ (dashed line),
and $A = 1$ (dashed-dotted line): (a) Coherence intensity;
(b) Spin polarization.
}
\label{fig:Fig.2}
\end{figure}

\newpage

%Figure 3
\begin{figure}[ht]
\vspace{9pt}
\centerline{
\hbox{ \includegraphics[width=7.5cm]{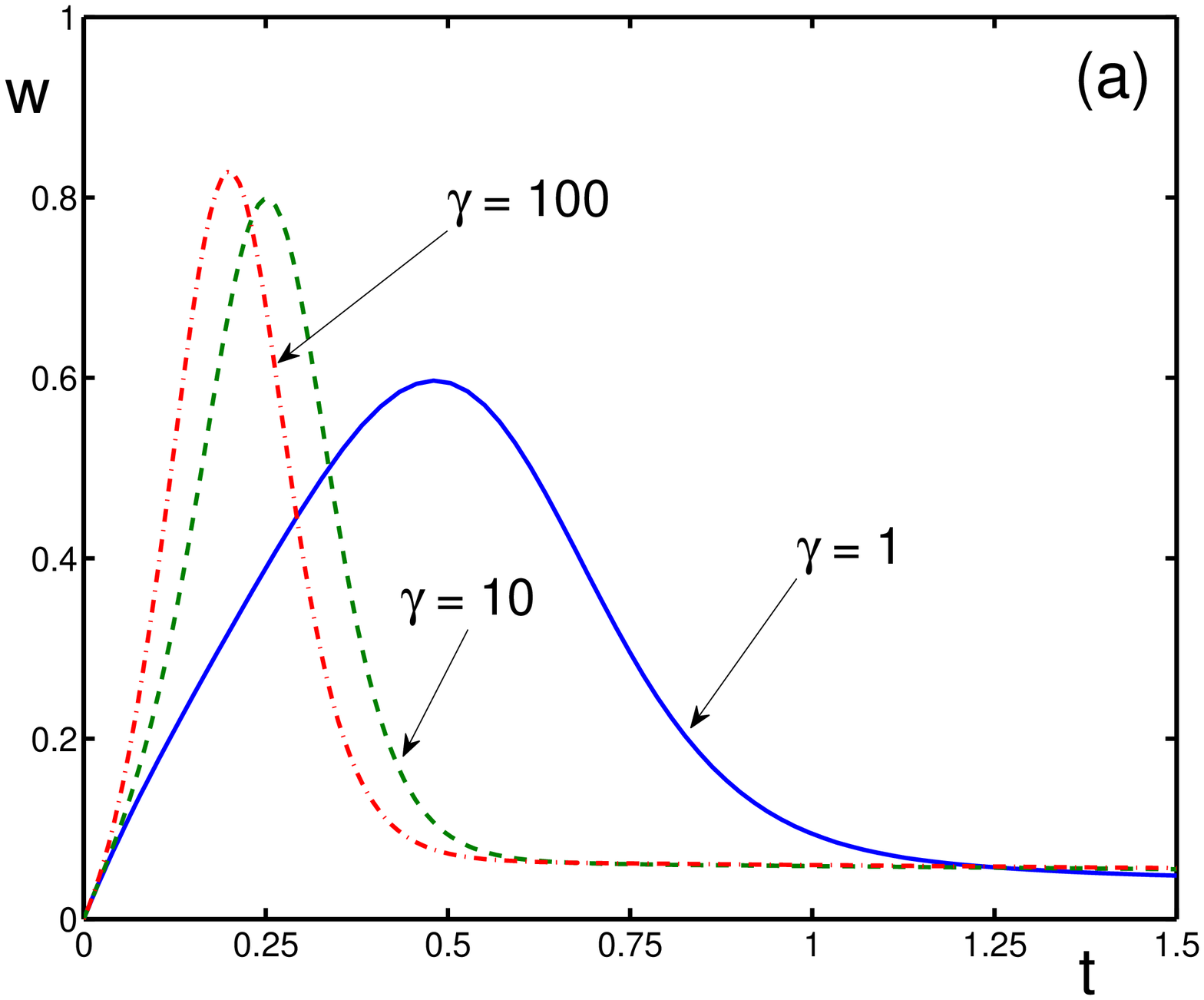} \hspace{2cm}
\includegraphics[width=7.5cm]{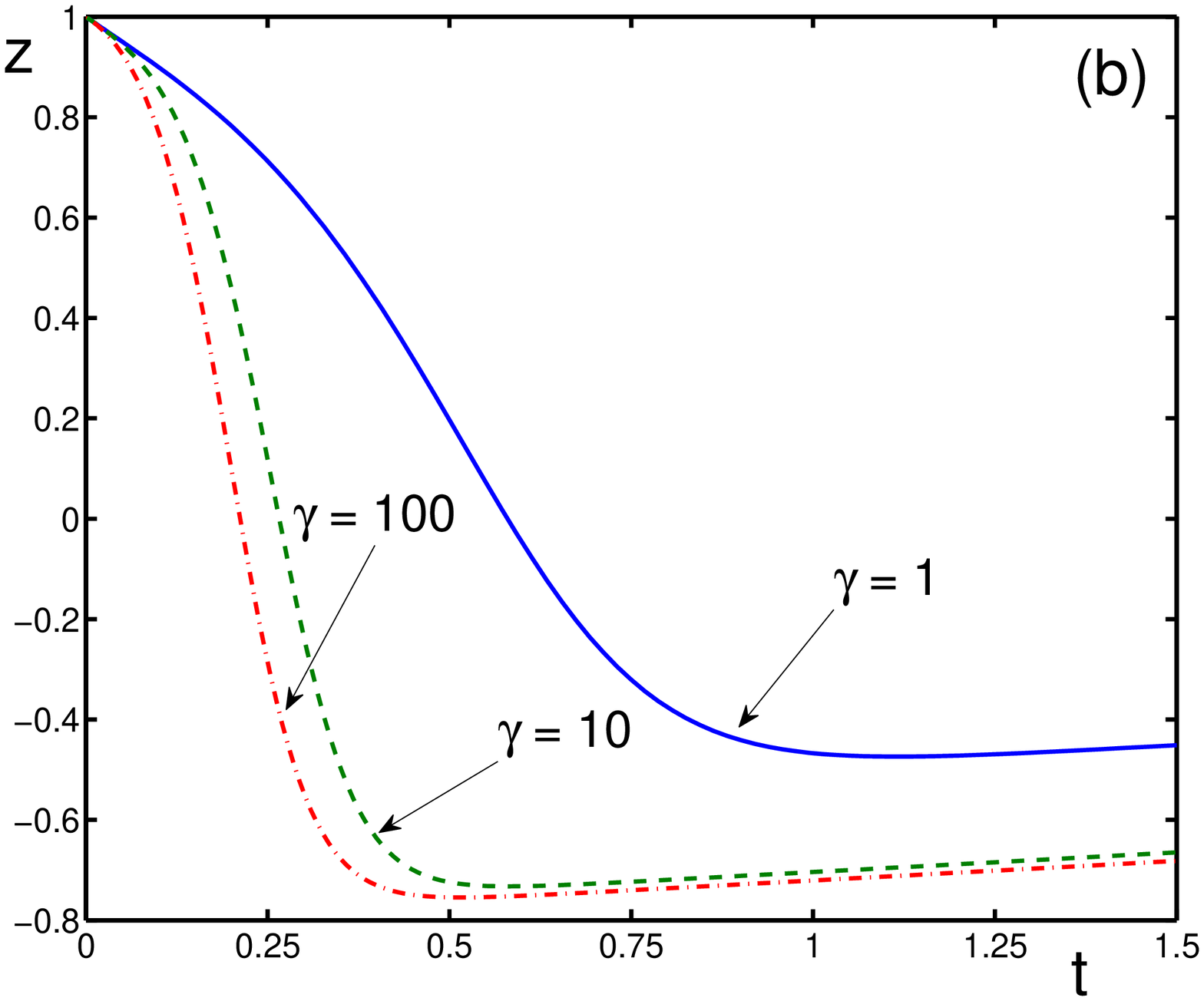} } }
\caption{Role of resonator attenuation for the parameters
$g = 10$, $A = 0.1$, and $\omega = 10$, under different attenuation
parameters, $\gamma = 1$ (solid line), $\gamma = 10$ (dashed line),
and $\gamma = 100$ (dashed-dotted line): (a) Coherence intensity;
(b) Spin polarization.
}
\label{fig:Fig.3}
\end{figure}

\vskip 3cm

%Figure 4
\begin{figure}[ht]
\vspace{9pt}
\centerline{
\hbox{ \includegraphics[width=7.5cm]{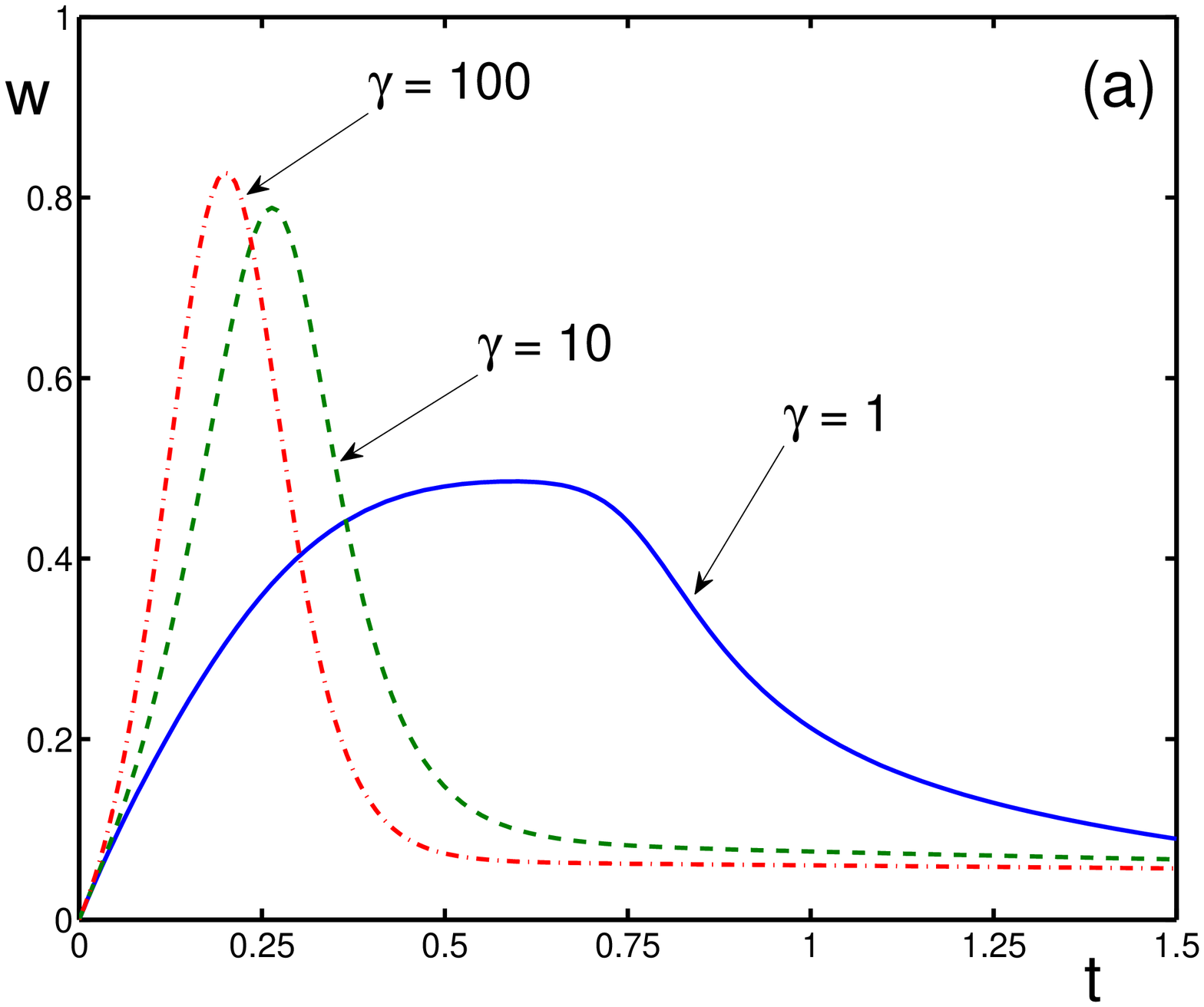} \hspace{2cm}
\includegraphics[width=7.5cm]{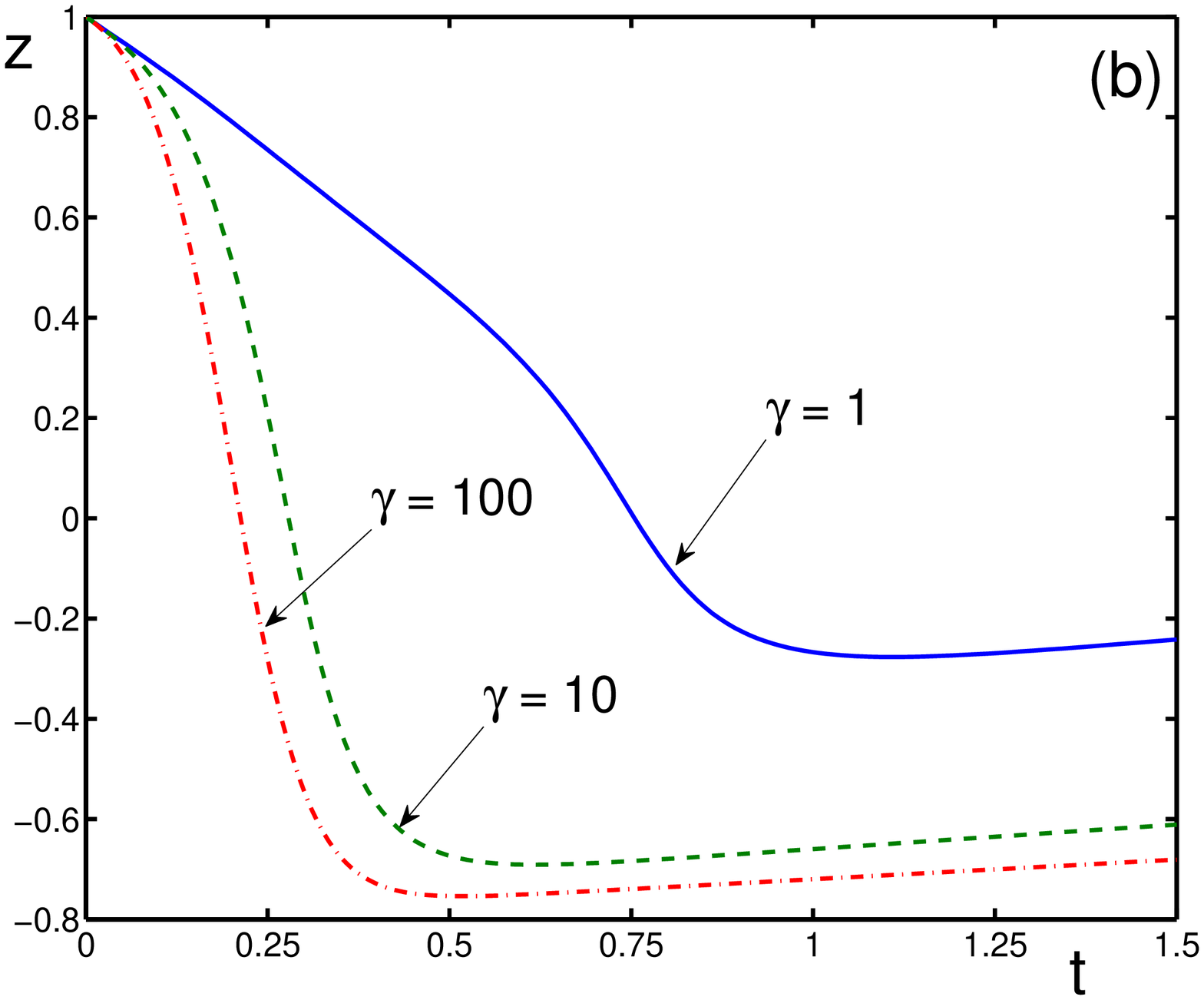} } }
\caption{Role of resonator attenuation for the parameters
$g = 10$, $A = 0.1$, but $\omega = 100$, under different attenuation
parameters, $\gamma = 1$ (solid line), $\gamma = 10$ (dashed line),
and $\gamma = 100$ (dashed-dotted line): (a) Coherence intensity;
(b) Spin polarization.
}
\label{fig:Fig.4}
\end{figure}

\newpage

%Figure 5
\begin{figure}[ht]
\vspace{9pt}
\centerline{
\hbox{ \includegraphics[width=7.5cm]{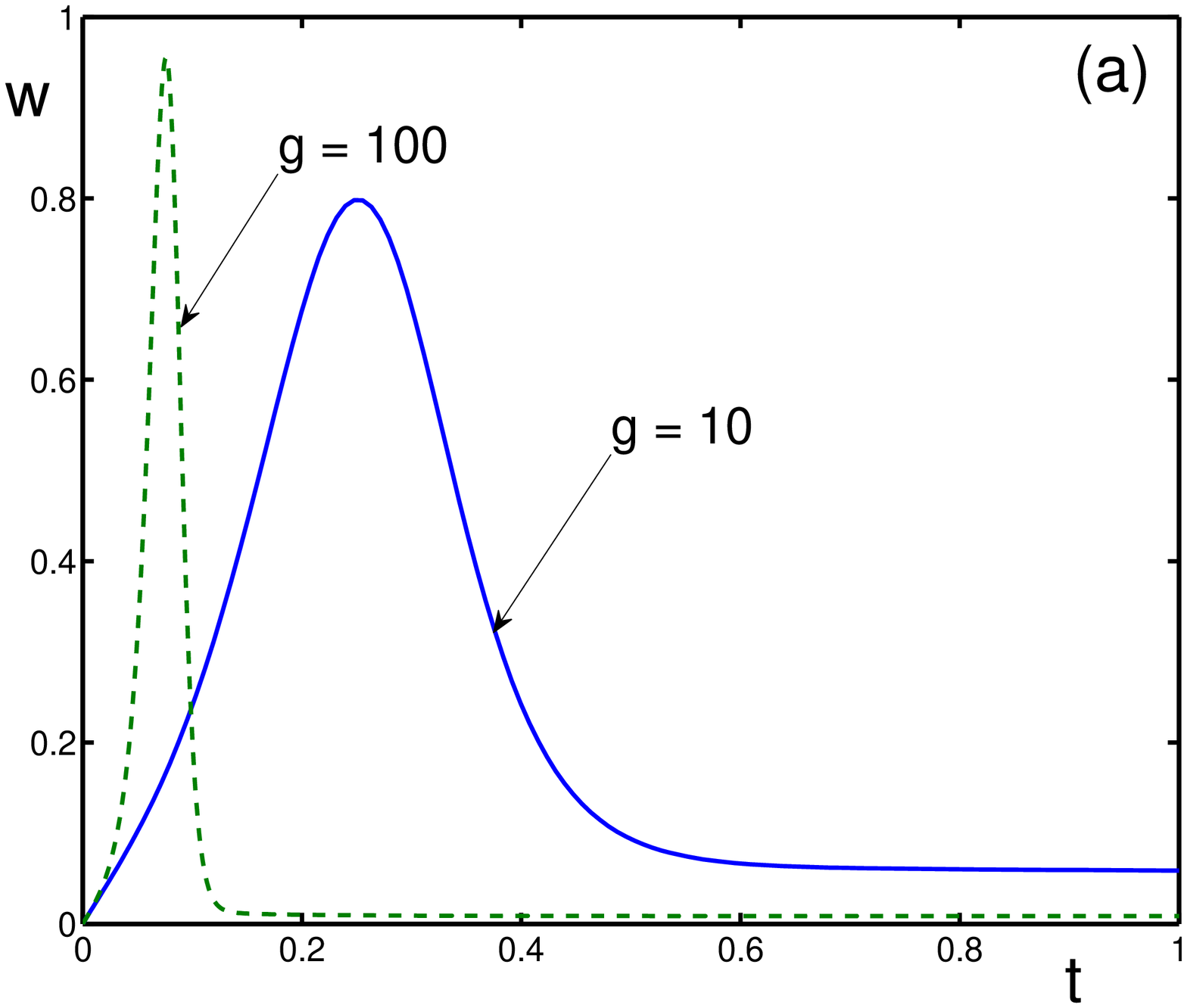} \hspace{2cm}
\includegraphics[width=7.5cm]{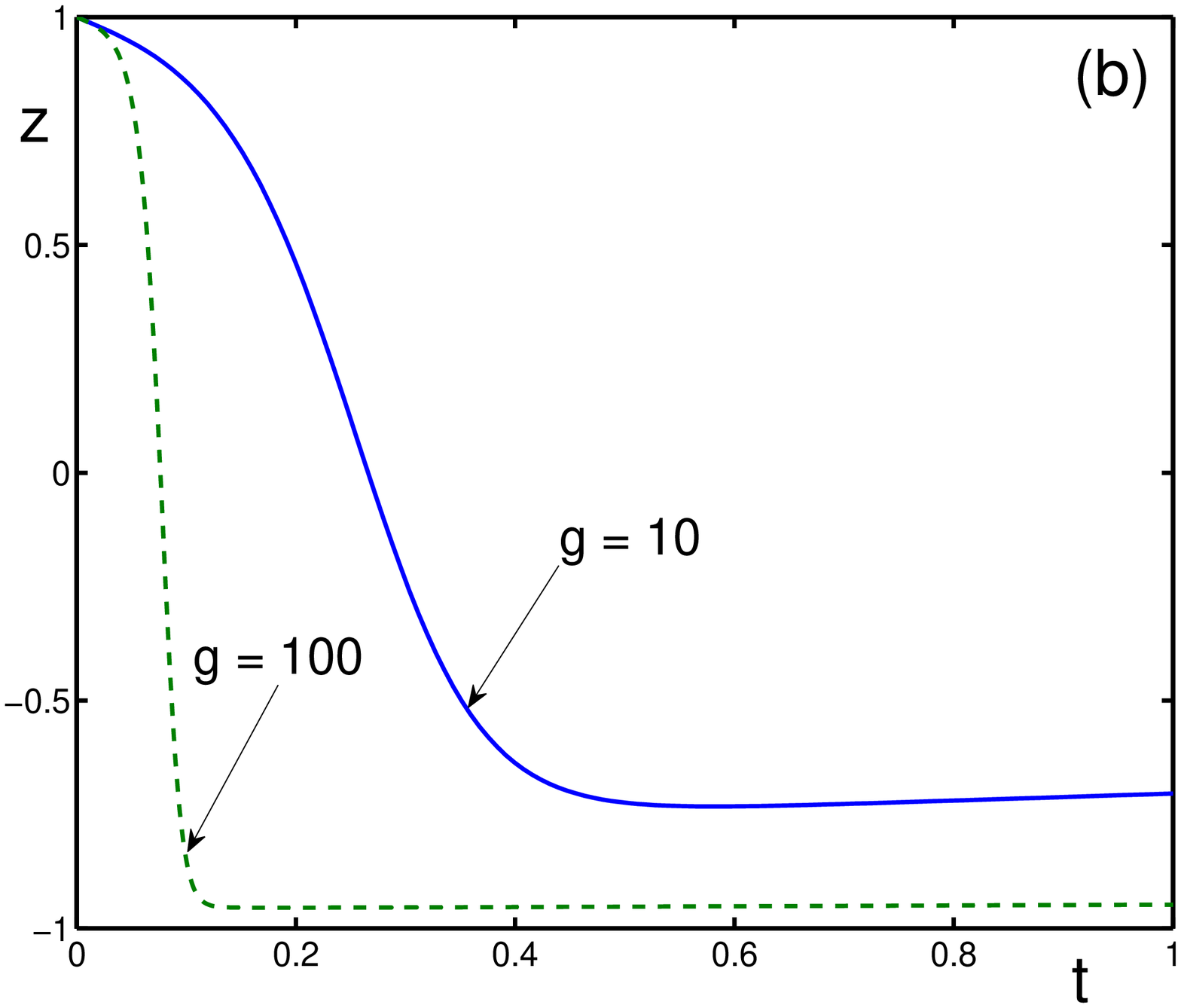} } }
\caption{Role of magnet-resonator coupling for the parameters
$\gamma = 10$, $A = 0.1$, and $\omega = 10$, under different coupling
parameters $g = 10$ (solid line) and $g = 100$ (dashed line):
(a) Coherence intensity; (b) Spin polarization.
}
\label{fig:Fig.5}
\end{figure}

\vskip 3cm

%Figure 6
\begin{figure}[ht]
\vspace{9pt}
\centerline{
\hbox{ \includegraphics[width=7.5cm]{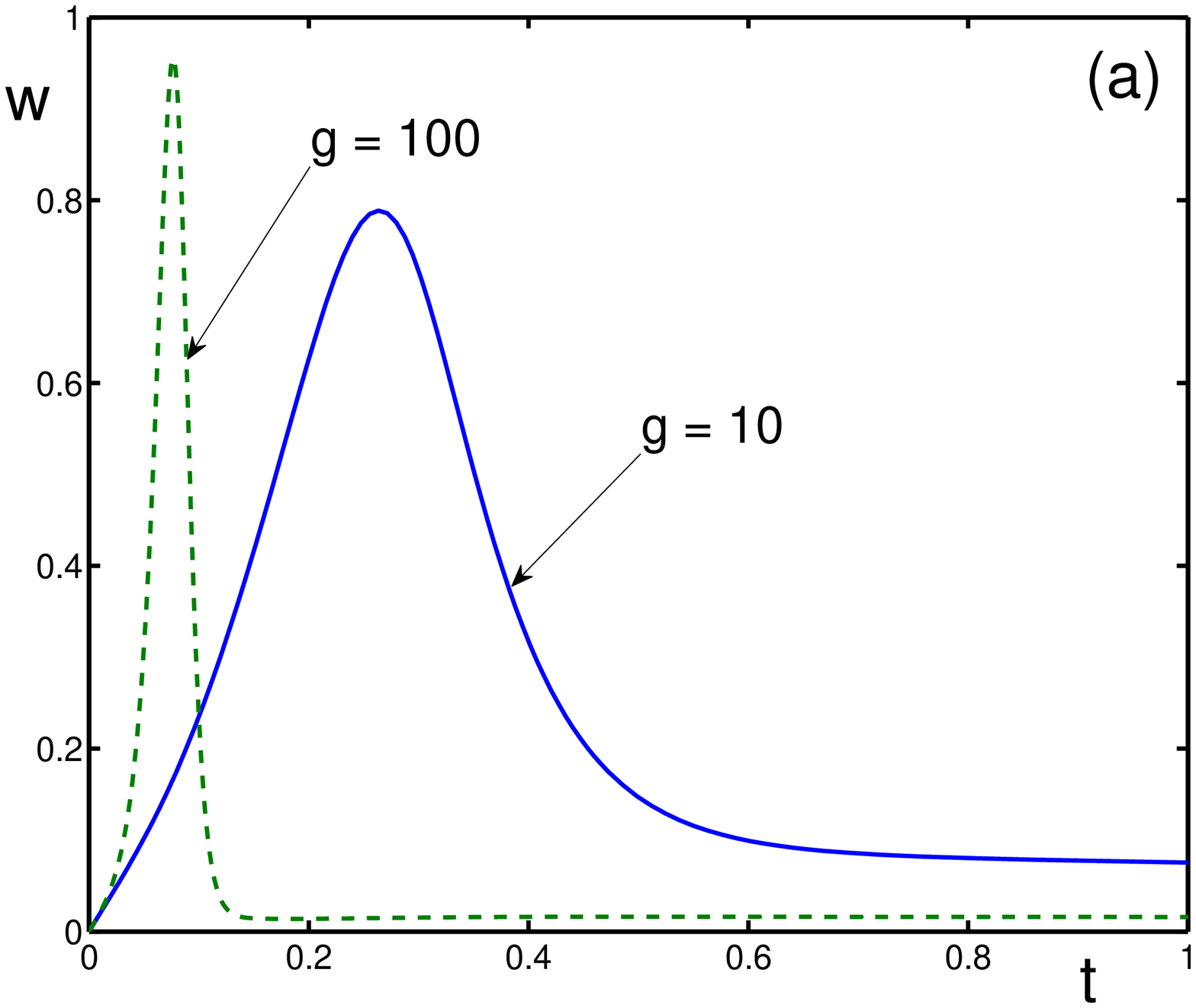} \hspace{2cm}
\includegraphics[width=7.5cm]{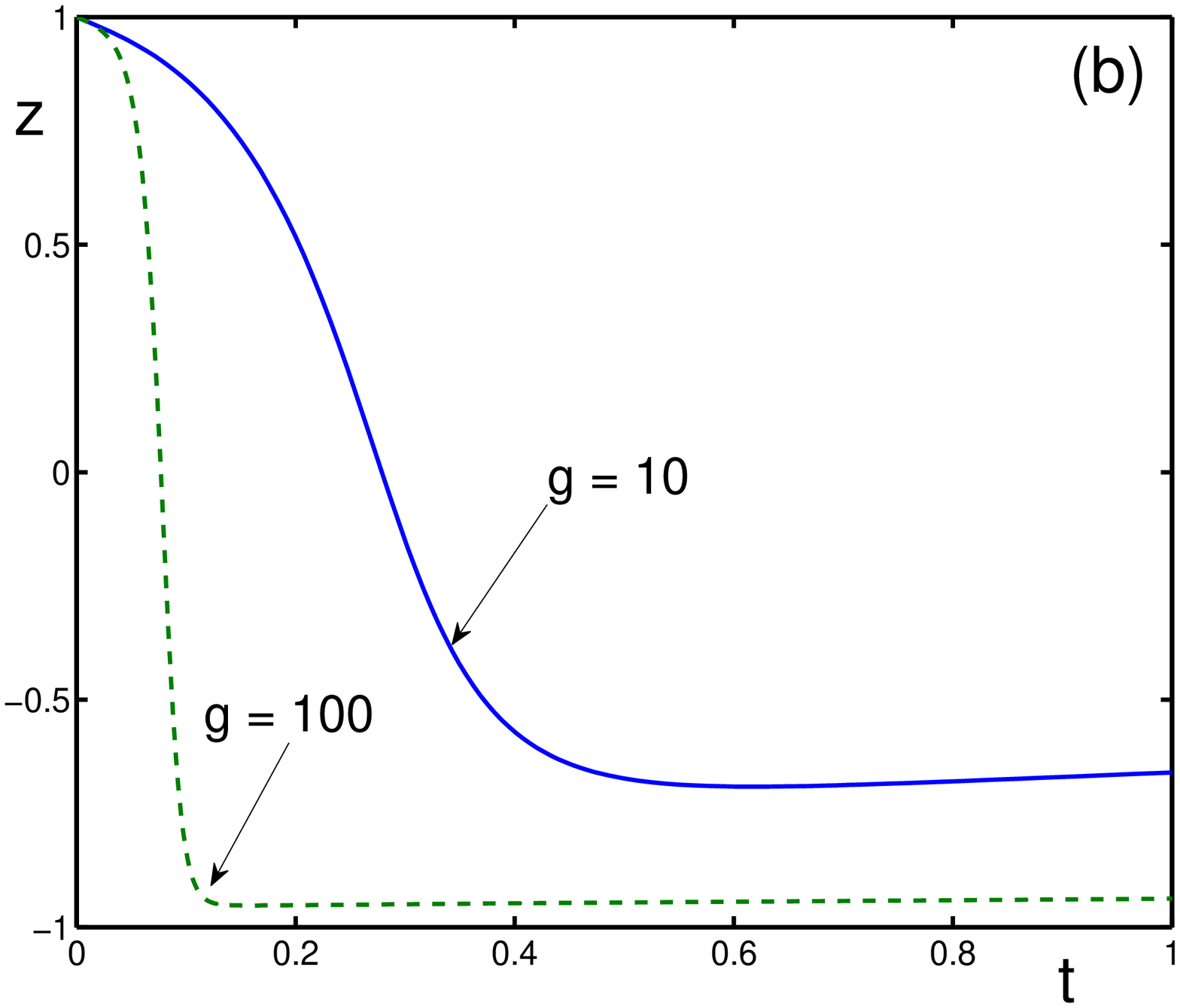} } }
\caption{Role of magnet-resonator coupling for the parameters
$\gamma = 10$, $A = 0.1$, but $\omega = 100$, under different coupling
parameters, $g = 10$ (solid line) and $g = 100$ (dashed line):
(a) Coherence intensity; (b) Spin polarization.
}
\label{fig:Fig.6}
\end{figure}

\newpage

%Figure 7
\begin{figure}[ht!]
\vspace{9pt}
\centerline{
\hbox{ \includegraphics[width=7.5cm]{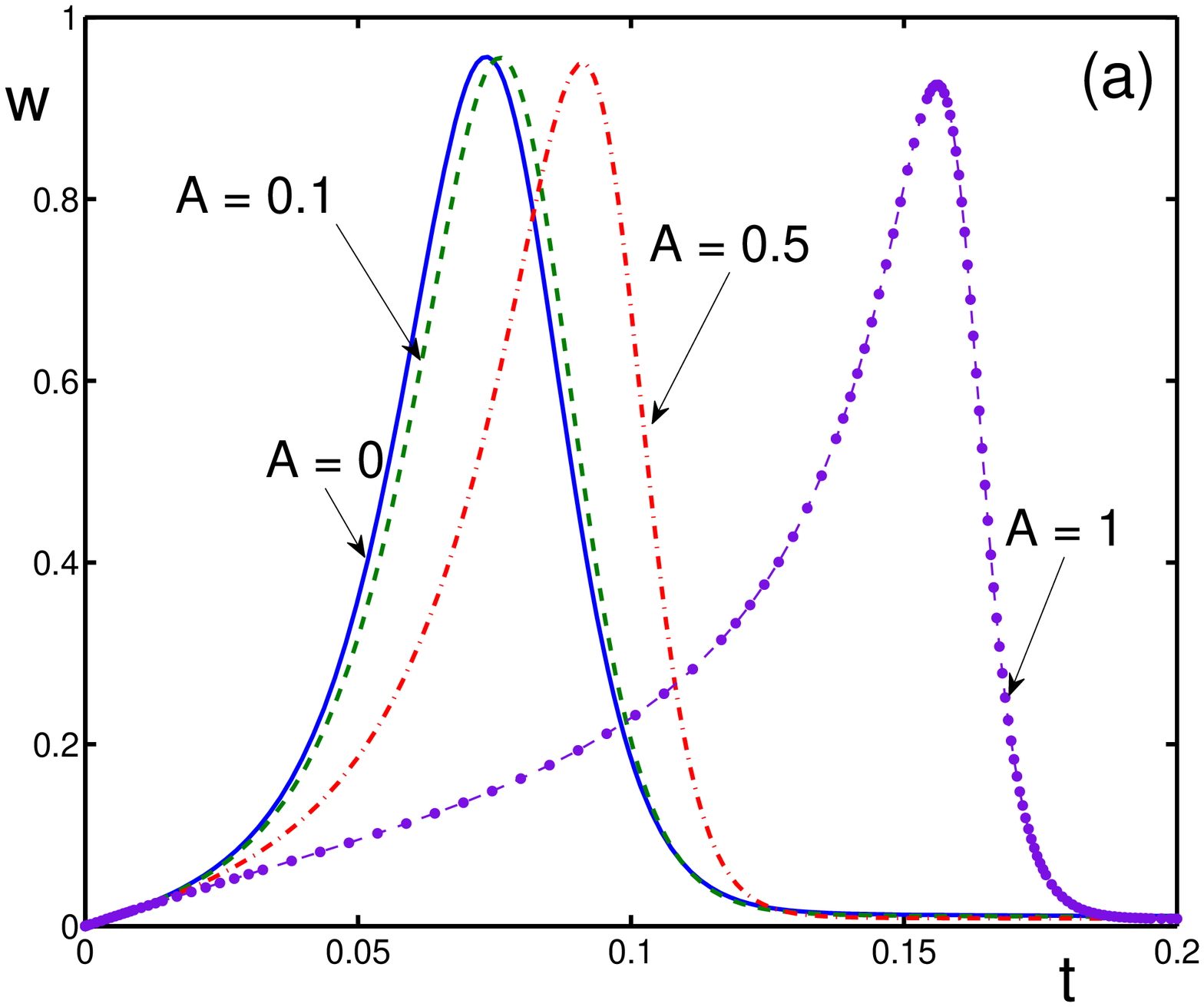} \hspace{2cm}
\includegraphics[width=7.5cm]{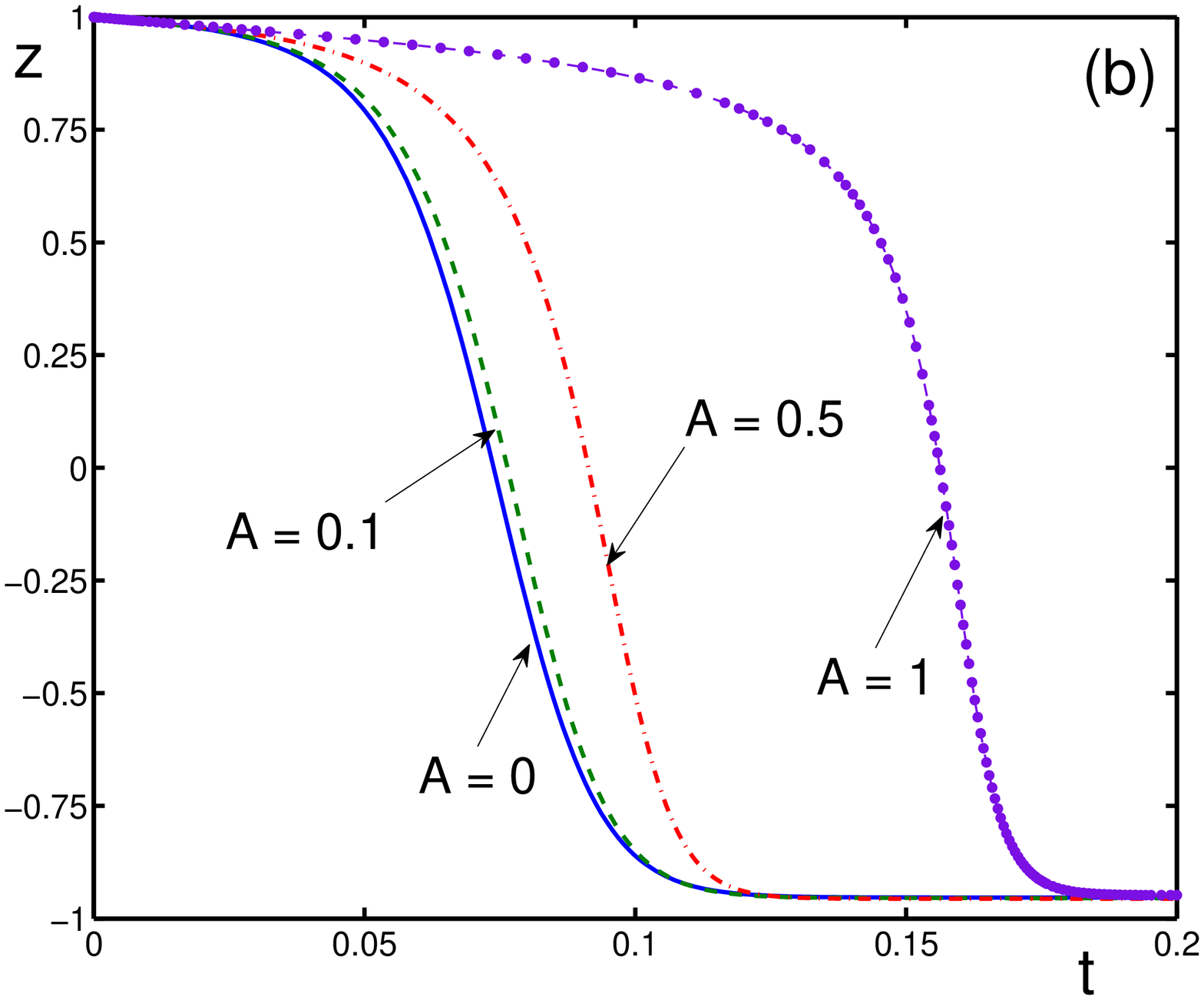} } }
\caption{Role of magnetic anisotropy for the parameters $\gamma = 10$,
$g = 100$, and $\omega = 10$, under different anisotropy parameters,
$A = 0$ (solid line), $A = 0.1$ (dashed line), $A = 0.5$
(dashed-dotted line), and $A = 1$ (dashed line with dots): (a) Coherence
intensity; (b) Spin polarization.
}
\label{fig:Fig.7}
\end{figure}

\vskip 3cm

%Figure 8
\begin{figure}[ht!]
\vspace{9pt}
\centerline{
\hbox{ \includegraphics[width=7.5cm]{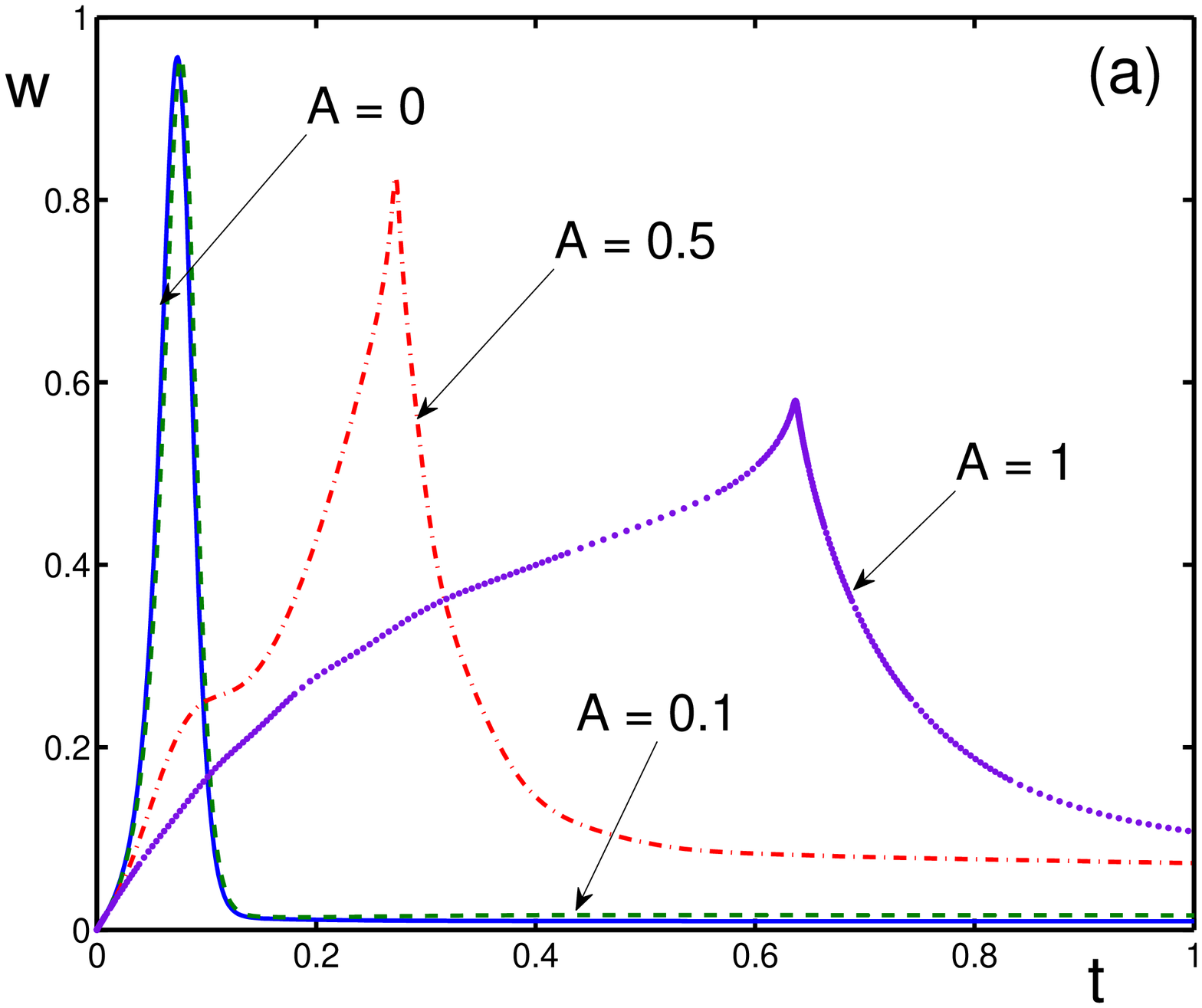} \hspace{2cm}
\includegraphics[width=7.5cm]{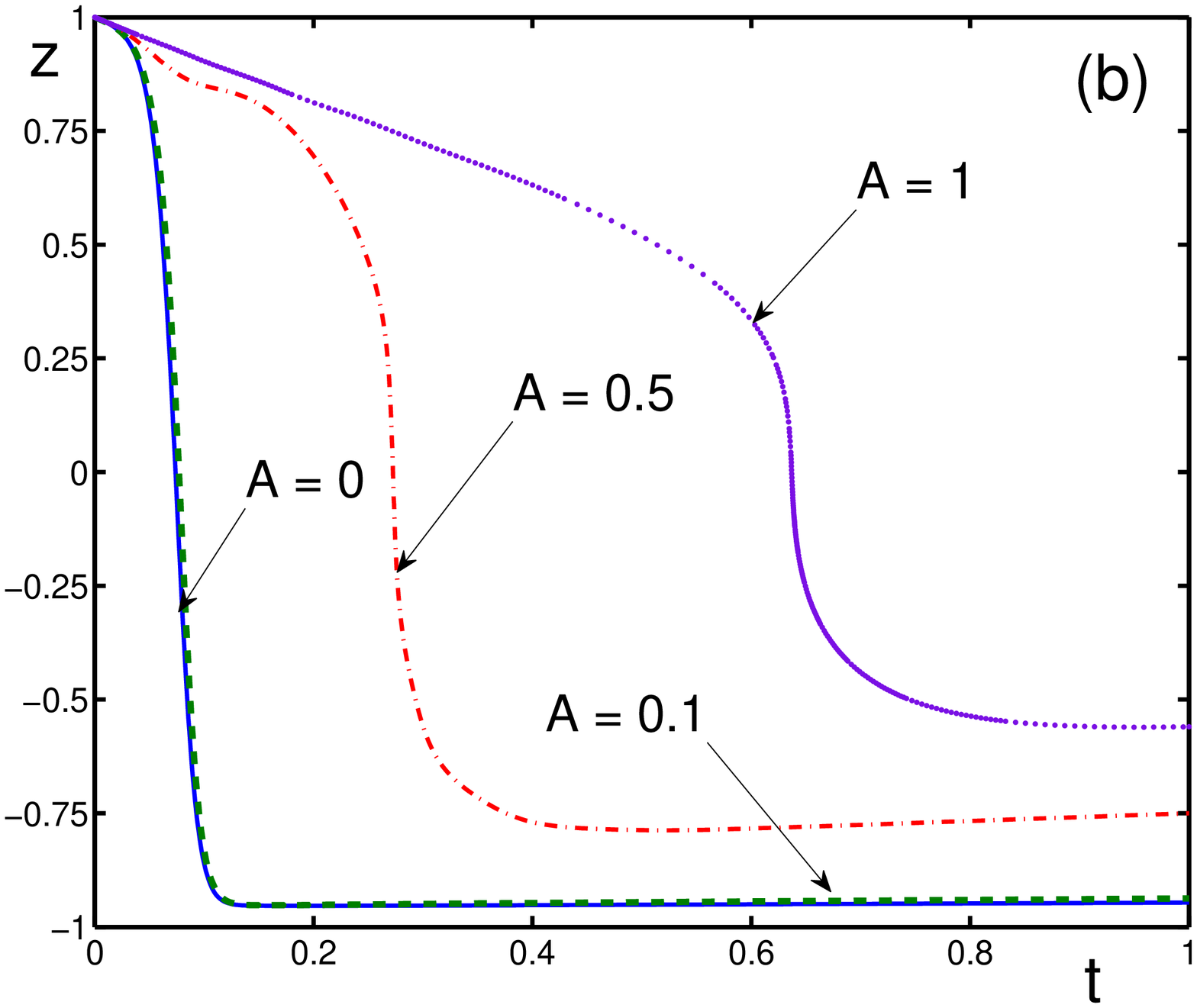} } }
\caption{Role of magnetic anisotropy for the parameters $\gamma = 10$,
$g = 100$, but $\omega = 100$, under different anisotropy parameters,
$A = 0$ (solid line), $A = 0.1$ ( dashed line), $A = 0.5$
(dashed-dotted line), and $A = 1$ (dotted line): (a) Coherence
intensity; (b) Spin polarization.
}
\label{fig:Fig.8}
\end{figure}

\end{document}